\documentclass[hyper]{JHEP3} % 10pt is ignored!

\usepackage{epsfig,multicol,bbm,amsmath}

\newcommand\fverb{\setbox\fverbbox=\hbox\bgroup\verb}
\newcommand\fverbdo{\egroup\medskip\noindent%
			\fbox{\unhbox\fverbbox}\ }
\newcommand\fverbit{\egroup\item[\fbox{\unhbox\fverbbox}]}
\newbox\fverbbox

\include{symbolsJHEP}

\def \all{A_{0L}}
\def \alr{A_{0R}}
\def \apl{A_{\| L}}
\def \apr{A_{\| R}}
\def \appl{A_{\bot L}}
\def \appr{A_{\bot R}}
\def \al{A_0}
\def \ap{A_{\|}}
\def \app{{A}_{\bot}}
\def \Aim{\ensuremath{A_{\mathrm{im}}}\xspace}
\def \AT#1{\ensuremath{A_T^{\left(#1\right)}}\xspace}
\def \FL{\ensuremath{F_{\mathrm{L}}}\xspace}
\def \zerocrossing{\ensuremath{q^2_0}\xspace}

\def \fig#1{Fig.~\ref{#1}}

\def \thetaL{\ensuremath{\theta_l}\xspace}
\def \thetaK{\ensuremath{\theta_\kaon}\xspace}

\def \Imag#1{\ensuremath{\mathrm{Im}(#1)}}

\def \be{\begin{equation}}
\def \ee{\end{equation}}
\def \bea{\begin{eqnarray}}
\def \eea{\end{eqnarray}}
\def \ben{\begin{enumerate}}
\def \een{\end{enumerate}}
\def \Bbar{\overline{\kern -0.24em B}}

\def \all{A_{0L}}
\def \alr{A_{0R}}
\def \apl{A_{\| L}}
\def \apr{A_{\| R}}
\def \appl{A_{\bot L}}
\def \appr{A_{\bot R}}
\def \al{A_0}
\def \ap{A_{\|}}
\def \app{{A}_{\bot}}
\def \mhat{\hat{m}_\kstar}

\def \kstar{{K^*}}

\def \sh{\hat{s}}
\def \Im{{\text{Im}}}
\def \Re{{\text{Re}}}

\def \braket#1#2#3{\langle #1|#2| #3\rangle}

\def \eq#1{Eq.~(\ref{#1})}

\def \fig#1{Fig.~\ref{#1}}

\def \nnu{\nonumber}

\def \ol#1{\overline{#1}}

\def \rf{Ref.~\cite}

\def \sec#1{Sec.~\ref{#1}}

\def \a{\alpha}

\def \g{\gamma}

\def \m{\mu}

\title{New observables in the decay mode $\boldmath{\BdbKsll}$}

\author{Ulrik Egede \\
	Imperial College London, London SW7~2AZ, United Kingdom\\
	E-mail: \email{U.Egede@imperial.ac.uk}}
\author{Tobias Hurth\\
	CERN, Dept. of Physics, Theory  Division, CH-1211 Geneva 23, Switzerland\thanks{Also SLAC, Stanford University, Stanford, CA 94309, USA.}\\
	E-mail: \email{Tobias.Hurth@cern.ch}}
\author{Joaquim Matias \\
	IFAE, Universitat Aut\`onoma de Barcelona, 08193 Bellaterra, 
Barcelona, Spain\\
	E-mail: \email{matias@ifae.es}}
\author{Marc Ramon\\
	IFAE, Universitat Aut\`onoma de Barcelona, 08193 Bellaterra, 
Barcelona, Spain\\
	E-mail: \email{mramon@ifae.es}}
\author{Will Reece \\
	Imperial College London, London SW7~2AZ, United Kingdom\\
	E-mail: \email{w.reece06@imperial.ac.uk}}

\preprint{
  \arXivid{0807.2589}\\
  UAB-FT-2008-652\\
  CERN-TH/2008-155\\
  SLAC-PUB-13307\\
}	% OR: \preprint{Aaaa/Mm/Yy\\Aaa-aa/Nnnnnn}
\abstract{We discuss the large set of observables available from the angular
  distributions of the decay \BdbKsll. We present a NLO analysis of all
  observables based on the QCD factorization approach in the low-dilepton mass
  region and an estimate of $\Lambda/m_b$ corrections. Moreover, we discuss
  their sensitivity to new physics. We explore the experimental sensitivities
  at \lhcb (10~\invfb) and \superlhcb (100~\invfb) based on a full-angular fit
  method and explore the sensitivity to right handed currents. We also show
  that the previously discussed transversity amplitude \AT{1} cannot be
  measured at the \lhcb experiment or at future $B$ factory experiments as it
  requires a measurement of the spin of the final state particles.}


\begin{document} 

\section{Introduction}
\label{sec:intro}
A major aim of particle physics in the LHC era is the discovery of new degrees
of freedom at the \tev energy scale which might contribute to our
understanding of the origin of electroweak symmetry breaking.  Rare $B$ and
kaon decays (for reviews see \cite{Buchalla:2008jp,Hurth:2003vb,Hurth:2007xa})
representing loop-induced processes are highly sensitive probes for new
degrees of freedom beyond the Standard Model~(SM) and will be used when making
indirect searches for these unknown effects.  It is well-known that the
indirect constraints on new physics~(NP) from the present flavour data
indicate a NP scale much higher than the electroweak scale when such new
effects are naturally parameterised by higher-dimensional operators. Thus, if
there is NP at the electroweak scale, then its flavour structure has to be
highly non-trivial and the experimental measurement of flavour-violating
couplings is mandatory. This `flavour problem', namely why flavour-changing
neutral currents are suppressed, has to be solved by any NP scenario at the
electroweak scale.

In this article we discuss theoretical and experimental preparations for an
indirect NP search using the rare decay~\BdbKsmm.  This exclusive decay was
first observed at Belle \cite{Ishikawa:2003cp}.  It offers a rich
phenomenology of various kinematical distributions beyond the measurement of
the branching ratio.  Some experimental analyses of those angular
distributions are already presented by the $B$ factories
\cite{Aubert:2003cm,Ishikawa:2006fh,Aubert:2006vb,Aubert:2008ju,Adachi:2008sk} but only the
large increase in statistics at
\lhcb~\cite{Dickens:2007ka,Dickens:2007zz,Egede:2007zz} for \BdbKsmm will make
much higher precision measurements possible. There are also great
opportunities at the future (Super-)$B$ factories in this
respect~\cite{Browder:2007gg,Bona:2007qt,Hewett:2004tv,Akeroyd:2004mj}. A
careful choice of observables needs to be made to take full advantage of this
exclusive decay as only in certain ratios such as \CP and forward-backward
asymmetries, the hadronic uncertainties cancel out in specific observables
making such ratios the only observables that are sensitive to NP. In this
respect the by now standard theoretical tools like QCD factorization (QCDf)~\cite{Beneke:2001at}
and its quantum field theoretical formulation, soft-collinear effective theory
(SCET), are crucial. They imply form factor relations which simplify the
theoretical structure of various kinematical distributions such that at least
at the leading order (LO) level any hadronic uncertainties cancel out. A
well-known example of this is the zero-crossing of the forward-backward
asymmetry. 

We construct new observables of this kind in the \BdbKsmm decay which have
very small theoretical uncertainties and good experimental
resolution. Moreover, it is possible to design the new observables for a
specific kind of NP operator within the model independent analysis using the
effective field theory approach.

Previously proposed angular distributions and \CP violating observables in
\BdbKsmm are reviewed in Ref.~\cite{Chen:2002bq,Hewett:2004tv}, and more
recently QCDf analyses of such angular distributions
\cite{Kruger:2005ep,Lunghi:2006hc} and \CP violating
observables~\cite{Bobeth:2008ij} were presented.

The paper is organised as follows: in Sec.~\ref{sec:distribution} we recall
the differential decay distribution in the \BdbKsmm decay; in
Sec.~\ref{sec:amplitudes} we recall the basic theoretical formulae which
are crucial for our construction of new observables; in
Sec.~\ref{sec:clean} we discuss the basic properties and symmetries of
potential observables and propose a new set of observables which are sensitive
to new right-handed currents and we also discuss the previously proposed
quantity $A_T^{(1)}$; in Sec.~\ref{sec:sensitivity} we explain our method
to calculate the experimental sensitivity obtainable with the statistics of
\lhcb to new and old observables; and finally in Sec.~\ref{sec:pheno} we
present our phenomenological analysis, in particular we analyse the
theoretical and experimental sensitivity to NP.  We also comment very
briefly on recent \babar\ measurements of certain angular distributions. In
appendices we make angular definitions explicit, provide the theoretical
framework for the derivation of the spin amplitudes, and present the
theoretical NLO expressions.

\section{Differential decay distribution}
\label{sec:distribution}

The decay \BdbKsll with $\Kstarzb \to \Km \pip$ on the mass shell, is
completely described by four independent kinematic variables, the lepton-pair
invariant mass squared, \qsq, and the three angles $\theta_l$, $\theta_{K}$,
$\phi$. Summing over the spins of the final particles, the differential decay
distribution of \BdbKsll can be written as
\begin{equation}
  \label{diff:four-fold}
  \frac{d^4\Gamma_{\Bdb}}{dq^2\,d\theta_l\, d\theta_K\, d\phi} = 
  \frac{9}{32 \pi} I(q^2, \theta_l, \theta_K, \phi) \sin\theta_l\sin\theta_K\:,
\end{equation}
with the physical region of phase space $4 m_l^2\leqslant q^2 \leqslant
(m_B-m_{\Kstar})^2$ and
\begin{eqnarray}
  \label{funcs:i}
  I &=& I_1 + I_2\cos 2\theta_l + I_3 \sin^2\theta_l\cos 2\phi + 
  I_4 \sin 2\theta_l \cos\phi + I_5 \sin\theta_l\cos\phi + \nonumber \\
  & &  + I_6 \cos\theta _l + I_7 \sin\theta_l\sin\phi + I_8 \sin 2\theta_l \sin\phi + I_9 \sin^2\theta_l\sin 2\phi.
\end{eqnarray}
The $I_i$ depend on products of the seven complex $K^*$ spin amplitudes,
$A_{\bot L/R}$, $A_{\| L/R}$, $A_{0 L/R}$, $A_t$ (see next section) with each of these a function of \qsq.  $A_t$ is
related to the time-like component of the virtual $K^*$, which does not
contribute in the case of massless leptons and can be neglected if the lepton
mass is small in comparison to the mass of the lepton pair.  We will consider
this case in our present analysis.  For $m_l = 0$, one finds
\cite{FK:etal,dmitri,CS:etal,4body:mass}:
\begin{subequations}
\label{eq:Isubis}
 \begin{eqnarray} 
    I_1 &=&  \frac{3}{4}\left(|\appl|^2 + |\apl|^2 + (L\to R)\right)
    \sin^2\theta_K + \left(|\all|^2 +|\alr|^2 \right) 
    \cos^2\theta_K \nonumber\\ 
    &\equiv& a \sin^2\theta_K+ b \cos^2\theta_K, \\
    I_2 &=&
    \frac{1}{4}( |\appl|^2+
    |\apl|^2)\sin^2\theta_K- |\all|^2\cos^2\theta_K + (L\to
    R)\nonumber \\ 
    &\equiv& c \sin^2\theta_K+ d \cos^2\theta_K , \\
    I_3 & = &\frac{1}{2}\bigg[(|\appl|^2 - |\apl|^2 )\sin^2\theta_K +
    (L\to R)\bigg]\equiv e \sin^2\theta_K, \\
    I_4 & = &
    \frac{1}{\sqrt{2}}\bigg[\Re
    (\all^{}\apl^*) \sin 2\theta_K + (L\to R)\bigg]\equiv f \sin
    2\theta_K,  \\
    I_5 & = &\sqrt{2}\bigg[\Re(\all^{}\appl^*) \sin2\theta_K -
    (L\to R)\bigg]\equiv g \sin 2\theta_K, \\
    I_6 & = &
    2\bigg[\Re
    (\apl^{}\appl^*) \sin^2\theta_K - (L\to R)\bigg]\equiv h
    \sin^2\theta_K, \\
    I_7 & = & \sqrt{2} \bigg[\Im (\all^{}\apl^*) \sin2\theta_K -
    (L\to R)\bigg]\equiv j \sin 2\theta_K, \\
    I_8 & = &
    \frac{1}{\sqrt{2}}\bigg[\Im
    (\all^{}\appl^*) \sin2\theta_K + (L\to R)\bigg]\equiv k \sin
    2\theta_K, \\
    I_9 & = & \bigg[\Im (\apl^{*}\appl) \sin^2\theta_K +
    (L\to R)\bigg]\equiv m \sin^2\theta_K. 
  \end{eqnarray}
\end{subequations}
The exact equations presented here depend on the definition of the angles 
which we for this reason have made explicit in Appendix~\ref{App:Kinematics}.

From comparing the amplitude terms in Eq.~(\ref{eq:Isubis}), we see that
$a=3c$ and $b=-d$ thus leaving nine independent parameters which can be fixed
experimentally in a full angular fit. Assuming massless leptons in the
theory we have on the other hand 12 parameters from the six complex \Kstarzb
spin amplitudes, $A_{\bot L/R}$, $A_{\| L/R}$, $A_{0 L/R}$. See
Sec.~\ref{sec:clean} for an analysis of the apparent mismatch between the 9 and 12 parameters.

\section{$K^*$ spin amplitudes}
\label{sec:amplitudes}
The six complex $K^*$ spin amplitudes 
under the assumption of massless leptons are related to the well-known helicity amplitudes 
(used for example in \cite{dmitri,CS:etal,ali:safir})  through 
\be\label{hel:trans} 
A_{\bot,\|} = (H_{+1}\mp H_{-1})/\sqrt{2}\, , \qquad A_0=H_0.
\ee
The amplitudes describe the $B\to K\pi$ transition 
and can be parameterised in terms of the seven $B\to K^*$  
form factors by means of a narrow-width approximation. They also depend 
on the  short-distance Wilson coefficients $C_i$ corresponding to the various
operators of the effective electroweak Hamiltonian. The precise definitions of 
the form factors and of the effective operators  are given in Appendix~\ref{sec:theory}.
One obtains~\cite{Kruger:2005ep}
\be\label{a_perp}
A_{\bot L,R}=N \sqrt{2} \lambda^{1/2}\bigg[
(\Ceff{9}\mp\C{10})\frac{V(q^2)}{m_B +m_\kstar}+\frac{2m_b}{\qsq} (\Ceff{7} + \Cpeff{7}) 
T_1(\qsq)\bigg], 
\ee
\be\label{a_par}
A_{\| L,R}= - N \sqrt{2}(m_B^2- m_\kstar^2)\bigg[(\Ceff{9}\mp \C{10})
\frac{A_1 (\qsq)}{m_B-m_\kstar} 
+\frac{2 m_b}{\qsq} (\Ceff{7} - \Cpeff{7}) T_2(\qsq)\bigg],
\ee
\begin{eqnarray}\label{a_long}
A_{0L,R}&=&-\frac{N}{2m_\kstar\sqrt{\qsq}} \times \nnu\\
        & & \times \bigg[
(\Ceff{9}\mp \C{10})\bigg\{(m_B^2-m_\kstar^2 -\qsq)(m_B+m_\kstar)A_1(\qsq)
 - \nnu \\
& & \qquad\qquad\qquad\qquad -\lambda \frac{A_2(\qsq)}{m_B +m_\kstar}\bigg\} + \nnu\\
& &\quad\ + {2m_b}(\Ceff{7} - \Cpeff{7}) \bigg\{
 (m_B^2+3m_\kstar^2 -\qsq)T_2(\qsq) - \nnu \\
& & \qquad\qquad\qquad\qquad\qquad\quad -\frac{\lambda}{m_B^2-m_\kstar^2} T_3(\qsq)\bigg\}\bigg],
\end{eqnarray}
where
\begin{equation}
  \label{eq:Lambdadef}
  \lambda= m_B^4  + m_{K^*}^4 + q^4 - 2 (m_B^2 m_{K^*}^2+ m_{K^*}^2 \qsq  + m_B^2 \qsq)
\end{equation}
and 
\begin{equation}
N=\sqrt{\frac{G_F^2 \a^2}{3\cdot 2^{10}\pi^5 m_B^3}
|V_{tb}^{}V_{ts}^{\ast}|^2 \qsq \lambda^{1/2}
\sqrt{1-\frac{4 m_l^2}{\qsq}}}.
\end{equation}

The crucial theoretical input we use in  our analysis is 
the observation that  in the limit
where the initial hadron is heavy and the final meson has a large
energy \cite{large:energy:limit} the hadronic form factors can be 
expanded in the small ratios $\Lambda_{\mathrm{QCD}}/m_b$ and
$\Lambda_{\mathrm{QCD}}/E$, where $E$ is the energy of
the light meson.  Neglecting  corrections of order $1/m_b$ and $\alpha_s$, 
the seven a priori independent $B\to K^*$ form factors  
 reduce to two universal form factors $\xi_{\bot}$ and  $\xi_{\|}$
\cite{large:energy:limit,Beneke:2001wa}. 
These relations  can be strictly derived within the QCDf and SCET 
approach and are given  in the appendix. 
Using those simplifications  the spin  amplitudes 
at leading order in $1/m_b$ and $\alpha_s$ have a very
simple form:
\be\label{LEL:tranversity:perp}
A_{\bot L,R}=  \sqrt{2} N m_B(1- \sh)\bigg[
(\Ceff{9}\mp\C{10})+\frac{2\hat{m}_b}{\sh} (\Ceff{7} + \Cpeff{7}) 
\bigg]\xi_{\bot}(E_\kstar),
\ee 
\be\label{LEL:tranversity:par}
A_{\| L,R}= -\sqrt{2} N m_B (1-\sh)\bigg[(\Ceff{9}\mp \C{10}) 
+\frac{2\hat{m}_b}{\sh}(\Ceff{7} -\Cpeff{7}) \bigg]\xi_{\bot}(E_\kstar)\, ,
\ee 
\begin{equation}\label{LEL:tranversity:zero}
A_{0L,R}= -\frac{Nm_B }{2 \hat{m}_\kstar \sqrt{\sh}} (1-\sh)^2\bigg[(\Ceff{9}\mp \C{10}) + 2
\hat{m}_b (\Ceff{7} -\Cpeff{7}) \bigg]\xi_{\|}(E_\kstar)\, ,
\end{equation}
with $\sh =  \qsq/m_B^2$, $\hat{m}_i =  m_i/m_B$. Here we neglected  
terms of $O(\hat{m}_{K^*}^2)$. 

Some remarks are in order:
\begin{itemize}
\item The theoretical simplifications are restricted to the kinematic region
  in which the energy of the $K^*$ is of the order of the heavy quark mass,
  i.e.~$\qsq \ll m_B^2$. Moreover, the influences of very light resonances
  below 1\gevgev question the QCD factorization results in that region.
  Thus, we will confine our analysis of all observables to the dilepton mass
  in the range, $1\gevgev \leqslant q^2 \leqslant 6\gevgev$.
\item Within the SM, we recover the naive quark-model prediction of $A_{\bot}=
  - A_{\|}$ \cite{quark:model,soares} in the $m_B\to \infty$ and $E_\kstar \to
  \infty$ limit (equivalently $\mhat^2\to 0$).  In this case, the $s$ quark is
  produced in helicity $-{1}/{2}$ by weak interactions in the limit $m_s\to
  0$, which is not affected by strong interactions in the massless case
  \cite{Burdman:2000ku}. Thus, the strange quark combines with a light quark
  to form a $K^*$ with helicity either $-1$ or $0$ but not $+1$. Consequently,
  the SM predicts at quark level $H_{+1}=0$, and hence $A_{\bot}= - A_{\|}$
  [cf.~\eq{hel:trans}], which is revealed as $|H_{-1}|\gg |H_{+1}|$ (or
  $A_{\bot}\approx - A_{\|}$) at the hadron level.
\item As noted in Ref.~\cite{Kruger:2005ep}, the contributions of the
  chirality-flipped operators ${\mathcal{O}}_{9,10}^\prime
  ={\mathcal{O}}_{9,10} (P_L\to P_R)$ can be included in the above amplitudes
  by the replacements $\Ceff{9,10} \to \Ceff{9,10} + \Cpeff{9,10}$ in
  Eq.~(\ref{LEL:tranversity:perp}), $\Ceff{9,10} \to \Ceff{9,10} -
  \Cpeff{9,10}$ in Eqs.~(\ref{LEL:tranversity:par}) and
  (\ref{LEL:tranversity:zero}).  However, they play a sub-dominant role in our
  NP analysis presented here.
\item The symmetry breaking corrections of order $\alpha_s$ can be calculated
  in the QCDf/SCET approach.  Those NLO corrections are included in our
  numerical analysis following Ref.~\cite{Beneke:2001at}.  The corresponding
  formulae for the case $\Cpeff{7} \neq 0$ are given in
  Appendix~\ref{sec:NLO}.
\item In general we have no means to calculate $\Lambda/m_b$ corrections to
  the QCDf amplitudes so they are treated as unknown corrections. This leads
  to a large uncertainty of theoretical predictions based on the QCDf/SCET
  approach. However, in specific examples one can combine QCDf/SCET results
  with calculations based on the QCD sum rule approach in order to estimate
  the leading power corrections.

  To take into account the present situation, we introduce a set of extra
  parameters, one for each spin amplitude, to explore what the effect of a
  possible $\Lambda/m_b$ correction could be:
  \begin{equation}
    \label{lambdamb}
    A_{\bot,\|,0}=A_{\bot,\|,0}^0 \left(1+c_{\bot,\|,0} \right) 
  \end{equation}
  where the `0' superscript stands for the QCD NLO Factorization
  amplitude and $c_{\bot,\|,0}$ are taken to vary in a range $\pm
  10\%$ which corresponds to a naive dimensional estimate. For each observable we look at, each of the amplitudes were varied in turn leaving the others at their central value. All the variations were then added in quadrature.
  Furthermore, we also give our final predictions taking into account
  further improvements on the power corrections and varying the 
  independent parameters in a less conservative range of $\pm 5\%$. 
\end{itemize}

\section{Theoretically clean observables}
\label{sec:clean}

\subsection{General criteria}
\label{sec:criteria}
We recall again that 2 of the 11 measurable distribution functions
$a,b,\dots,m$ of the differential decay distribution in the limit $m_\ell^2
\ll q^2$, defined in Eq.~\ref{eq:Isubis}, include redundant information due to
the relations $a=3c$ and $b=-d$.  So in principle there are 9 independent
observables. However, the dependence of those functions on the six complex
theoretical spin amplitudes, $A_{\bot L/R}$, $A_{\| L/R}$ and $A_{0 L/R}$, is
special. By inspection one finds that the distribution functions are
{\it invariant} under the following three independent symmetry transformations
of the spin amplitudes: global phase transformation of the $L$-amplitudes
\begin{equation}
  \label{Sym1}  
  A^{'}_{\bot L} = e^{i \phi_L} A_{\bot L},\qquad  A^{'}_{\| L} = e^{i \phi_L} A_{\| L},\qquad   A^{'}_{0  L} = e^{i \phi_L} A_{0  L},
\end{equation}
global phase transformation of the $R$-amplitudes
\begin{equation}
  \label{Sym2}  
  A^{'}_{\bot R} = e^{i \phi_R} A_{\bot R},\qquad  A^{'}_{\| R} = e^{i \phi_R} A_{\| R},\qquad   A^{'}_{0  R} = e^{i \phi_R} A_{0  R},
\end{equation}
and a continuous $L \leftrightarrow R$ rotation
\begin{subequations}
  \label{Sym3}
  \begin{eqnarray}
    A^{'}_{\bot L} &=& + \cos\theta A_{\bot L}  +  \sin\theta  A^*_{\bot R}\\
    A^{'}_{\bot R} &=& -  \sin\theta A^*_{\bot L}  +  \cos\theta  A_{\bot R}\\
    A^{'}_{0 L} &=&  + \cos\theta A_{0 L}  - \sin\theta  A^*_{0 R}\\
    A^{'}_{0 R} &=& + \sin\theta A^*_{0 L}  +  \cos\theta  A_{0 R}\\
    A^{'}_{\| L} &=& + \cos\theta A_{\| L}  - \sin\theta  A^*_{\| R}\\
    A^{'}_{\| R} &=& + \sin\theta A^*_{\| L}  +  \cos\theta  A_{\| R}\, .
  \end{eqnarray}
\end{subequations}
Normally, there is the freedom to pick a single global phase, but as $L$ and
$R$ amplitudes do not interfere here, two phases can be chosen arbitrarily as
reflected in the first two transformations. The third symmetry reflects that
an average is made over the spin amplitudes to obtain the angular
distribution. So it is clear that only $9$ out of the $12$ parameters arising
from the 6 complex amplitudes are independent which fits exactly with the $9$
independent measurable distribution functions.

A consequence of the three symmetries is that any observable based on the
differential decay distribution has also to be invariant under the same
symmetry transformations.

Besides the mandatory criterion above there are further criteria required for
an interesting observable:
\begin{description}
\item[Simplicity:] A simple functional dependence on the 9 independent
  measurable distribution functions; at best it should depend only from one or
  two in the numerator and denominator of an asymmetry.
\item[Cleanliness:] At leading order in $\Lambda/m_b$ and in $\alpha_s$ the
  observable should be independent of any form factor, at best for all
  $q^2$. Also the influence of symmetry-breaking corrections at order
  $\alpha_s$ and at order $\Lambda/m_b$ should be minimal.
\item[Sensitivity:] The sensitivity to the $\Cpeff{7}$ Wilson coefficient
  representing NP with another chirality than in the SM should be maximal.
\item[Precision:] The experimental precision obtainable should be good enough
  to distinguish different NP models.
\end{description}

In the limit where the \Kstarzb meson has a large energy, only two independent
form factors occur in $A_{0 L/R}$ and in $A_{\bot L/R}$ and $A_{\|
  L/R}$. Clearly, any ratio of two of the nine measurable distribution functions
proportional to the same form factor fulfil the criterion of symmetry,
simplicity, and theoretical cleanliness up to $\Lambda/m_b$ and $\alpha_s$ 
corrections. However, the third criterion, a sensitivity to a special kind of
NP and the subsequent requirement of experimental precision, singles
out particular combinations. In this paper we focus on new right-handed
currents. Other NP sensitivities may single out other observables as
will be analysed in a forthcoming paper~\cite{secondpaper}.

\subsection{Observables}
\label{sec:observables}
There are some proposals for theoretical clean observables already in the 
literature which we should briefly discuss in view of the above criteria:
\begin{itemize}
\item The forward backward asymmetry is the most popular quantity in
  the \BdbKsmm decay~\cite{Ali:1991is}. In terms of the \Kstarzb spin
  amplitudes it can be written as~\cite{Beneke:2001at,Feldmann:2002iw}
  \begin{equation}
    \label{eq:AFBdef}
    \AFB = \frac{3}{2}\frac{\Re(\apl\appl^*) - \Re(\apr\appr^*)}
    {|\al|^2 + |\ap|^2 + |\app|^2}       
  \end{equation}
  where
  \begin{equation}
    A_i A^*_j\equiv A^{}_{i L}(q^2) A^*_{jL}(q^2)+ A^{}_{iR}(q^2) A^*_{jR}(q^2) 
    \quad 
    (i,j  = 0, \|, \perp)\, .
  \end{equation}
  While the criteria of symmetry and simplicity are fulfilled, the form factors
  cancel out only at the specific value of \qsq where $\AFB=0$. Thus the
  measurement provide only a single clean number, the zero crossing point,
  rather than a theoretically clean distribution. 
\item The fractions of the \Kstarzb polarisation
  \begin{eqnarray}\label{def:frac:pol}
    F_L(\qsq)
    & = & \frac{|{A}_0|^2}{|{ A}_0|^2 + |{A}_{\|}|^2
      + |A_\perp|^2} ,  \\
    F_T(\qsq) & = & 1 - F_L(\qsq) = \frac{|{A}_\bot|^2+|{A}_\||^2}{|{ A}_0|^2 + |{A}_{\|}|^2 + |A_\perp|^2}\, ,
  \end{eqnarray}
  and the $K^*$ polarisation parameter
  \begin{equation}
    \label{def:pol:param}
    \alpha_{K^*}(\qsq) = \frac{2 F_L}{F_T} - 1 = 
    \frac{2|{ A}_0|^2}{|{A}_{\|}|^2 + |{A}_{\perp}|^2}-1\, .
  \end{equation}
  All fulfil the criteria of symmetry and simplicity, but the form factors do
  not cancel in the LO approximation; thus, suffering from larger
  hadronic uncertainties. The fraction of the \Kstarb polarisation can be
  measured from the angular projections alone and the first experimental
  measurements of $F_L$ with limited accuracy are
  available~\cite{Aubert:2008ju,Adachi:2008sk}.
\item Defining the helicity distributions $\Gamma_\pm = |H^L_{\pm1}|^2 +
  |H^R_{\pm1}|^2$ one can construct~\cite{dmitri}
  \begin{equation}
    \label{eq:AT1Def}
    \AT{1}=\frac{\Gamma_{-}-\Gamma_{+}}{\Gamma_{-}+\Gamma_{+}}
    =\frac{-2\Re(\ap^{}\app^*)}{|\app|^2 + |\ap|^2} \, .
  \end{equation}

  It has been shown\cite{Kruger:2005ep,Lunghi:2006hc} that this quantity 
  has adequate cleanliness and is is very sensitive to right-handed currents,
  making an ideal observable if just these two criteria were
  sufficient. However, the quantity \AT{1} does not fulfil the most important
  criterion of symmetry.  The important consequences out of this observation
  are briefly discussed in the next subsection.
\item The other transversity amplitude, first proposed
  in~\cite{Kruger:2005ep}, is defined as
  \begin{equation}
    \label{eq:AT2Def}
    \AT{2} =\frac{|\app|^2 - |\ap|^2}{|\app|^2 + |\ap|^2}\, .
  \end{equation}
  It obviously fulfils all three criteria of symmetry, simplicity and
  theoretical cleanliness. It is also rather sensitivity to \Cpeff{7} as one
  can see by inspection of the LO formulae of the \Kstarzb amplitudes in
  Eqs.~(\ref{LEL:tranversity:perp}-\ref{LEL:tranversity:zero}); in this
  approximation it is directly proportional to \Cpeff{7}, thus vanishes in the
  SM.
\end{itemize}
By inspection of the formulae of the $K^*$ spin amplitudes in terms of the
Wilson coefficients and the SCET form factors at the LO approximation,
Eqs.~(\ref{LEL:tranversity:perp}-\ref{LEL:tranversity:zero}),
one is led to some  new observables which fulfil the first
three criteria {\it and} have an enhanced sensitivity to \Cpeff{7}. They are
defined as
\begin{equation}
  \label{eq:AT3Def}
  \AT{3} =
  \frac{|\all\apl^* + \alr^*\apr|}{\sqrt{ |\al|^2  |\app|^2}}\, ,
\end{equation}
and
\begin{equation}
  \label{eq:AT4Def}
  \AT{4} = \frac{|A_{0L} A_{\perp L}^* - A_{0R}^* A_{\perp R}| }{|
    A_{0L}^* A_{\|L}+A_{0R} A_{\|R}^*|}\, ,
\end{equation}
One could also consider the real and imaginary parts of \AT{3}.

There are no further independent quantities which fulfil the criteria we have
set out.  However, when we will consider NP sensitivities beyond \Cpeff{7}
further observables may be singled out~\cite{secondpaper}. 

\subsection{The problem with \boldmath{\AT{1}}}
Contrary to the case of \AT{i} with $i=2,3,4$, it is not possible to extract
\AT{1} from the full angular distribution.  This is a direct consequence of
the fact that the quantity \AT{1} is not invariant under the
symmetry~(\ref{Sym3}) of the distribution function~(\ref{diff:four-fold})
which represent the complete set of observables in the case spins of the final
states are summed up. Let us elaborate further on this surprising observation;
it seems practically not possible to measure the helicity of the final states
on a \textit{event-by-event} basis. At the forthcoming \lhcb experiment for
example one only measures the charge, the three-momentum of the final state
particles and its nature through different types of particle identification.
So one has the four-momentum for each particle and its charge. The situation
does not look different for the present $B$ factories and their future
upgrades. While the $e^+e^-$ environment is much simpler there is still no
practical way to measure the spin of the muons on an \textit{event-by-event}
basis. We should emphasise that this is a practical and not a conceptual
problem; in a \textit{gedanken} experiment where the helicity of the
individual final state leptons are measured, it would indeed be possible to
measure \AT{1}. So while \AT{1} is in principle a good observable, we cannot
see any way it can be measured at either \lhcb or at a Super-$B$ factory with electrons or muons in the final state.

\section{Method to calculate experimental sensitivity}
\label{sec:sensitivity}
In this section we explain how to investigate the sensitivity to the angular
observables presented in Sec.~\ref{sec:clean} using a toy Monte Carlo
model. We estimate the statistical uncertainty on all observables with
statistics corresponding to 5 years of nominal running at \lhcb (10\invfb) and
comment on the experimental prospects for a measurement at the end of an
upgrade to \lhcb (100\invfb). For the estimates here we are only considering
the final state with muons.

\subsection{\BdbKsll decay model}            
The angles \thetaL, \thetaK and $\phi$, as well as the \qsq of the lepton pair
can be measured with small uncertainty and no experimental resolution effects
need to be considered. A toy Monte Carlo model of the decay was created using
\eq{diff:four-fold} as a probability density function (PDF) normalised to
the width,
\begin{equation}\label{eqn:total_width}
\int_{q^2_{\min}}^{q^2_{\max}}\frac{d\Gamma}{dq^2}dq^2 \, .
\end{equation}
It is parameterised in terms of the real and imaginary parts of the spin
amplitudes where each of these amplitudes is \qsq dependent. A simple approach,
where the data is divided into regions of \qsq and the spin amplitudes
determined within these, will not work; the coefficients in front of the
different angular components as seen in \eq{diff:four-fold} depend in a
non-linear way on the spin amplitudes meaning that the angular distribution
after integration over a bin in \qsq cannot be expressed in terms of
\eq{diff:four-fold} with some \qsq-averaged spin amplitudes.  Instead an
approach is used where the \qsq dependence of each of the spin amplitudes is
parameterised as a function of \qsq.

A special choice of the symmetry transformations described in
Sec.~\ref{sec:clean} can be used to reduce the number of parameters. Here we
use the first two symmetry transformations Eqs.~(\ref{Sym1}) and (\ref{Sym2})
to get rid of the two phases in $A_{0 R}$ and $A_{0 L}$. Then the third
transformation \eq{Sym3} is used with $\theta = {\rm arctan}(- A_{0 R} /A_{0
  L})$ leading to $A_{0 L}$ being real and $A_{0 R}=0$ thus disappearing
completely from the parametrisation. At a given value of \qsq we are thus left
with 9 parameters corresponding to the real and imaginary components of
$A_{\|L,R}$ and $A_{\perp L,R}$ and the real component of $A_{0L}$. We now
parameterise each of these spin amplitudes as a $2^{\mathrm{nd}}$ order
polynomial. Through the polynomial ansatz we are introducing a weak model
dependence; we checked that the error introduced by this was significantly
smaller than the corresponding experimental errors across the squared dimuon
mass range $1\gevgev < \qsq < 6\gevgev$. To describe the full \qsq and angular
dependence of the decay we thus need 27 parameters. As a final step we
recognise that an absolute measurement of the total width is difficult to
obtain in a hadronic environment such as \lhcb and fix the value of $A_{0 L}$
to 1 at a reference value of \qsq thus reducing the number of free parameters
to 26. This last step has no influence on the experimental determination of
any of the observables discussed in this paper as they are all formed as
ratios where the total width cancels out. While no longer sensitive to the
absolute width we are still sensitive to the shape of the differential width
as a function of \qsq.

We follow the resolution, yield and background numbers
in~\cite{Dickens:2007ka} to construct a model that includes a realistic level
of background. The signal is assumed to have a Gaussian
distribution in $m_\B$ with a width of $14\mev$ in a window of $m_\B \pm
50\mev$ and a Breit-Wigner in $m_{K\pi}$ with width $48\mev$ in a window of
$m_{\Kstarz} \pm 100\mev$. A simplified background model is included; it is
flat in all angles, effectively treating all background as combinatorial, but
follows the \qsq distribution of the signal. Acceptance and \CP violation
effects are neglected allowing us to treat \BdbKsmm and its charge conjugate
simultaneously. We do not include any contributions from non--resonant
\mbox{$\Bdb \to \Km\pip\mup\mun$}.

Using the toy Monte Carlo model, a dataset for the observables \thetaL, \thetaK, $\phi$ and \qsq
can be generated with the calculated values of the spin amplitudes as input
without making use of the polynomial ansatz. Physics beyond the SM can be
included in a straightforward way by providing the relevant spin
amplitudes. Using the yield and background estimates
from~\cite{Dickens:2007ka} and assuming a flat efficiency for the signal as a
function of \qsq we use on average 4032 signal events and 1168 background
events in the \qsq interval from $4 m_\mu^2$ to $9\gevgev$ in a dataset of
2\invfb. These are scaled lineally in order to obtain 10\invfb and 100\invfb
yield estimates. For each dataset we generate, the signal and background
numbers are varied according to Poisson statistics.

The purpose of the toy Monte Carlo model is to enable us to illustrate the
methodology of this approach and be able to make precise statements on the
relative performance of a full angular fit compared to just looking at
projections. Accurate estimates of the resolution in each parameter will only
be possible with a complete detector simulation and with a complete
understanding of the actual detector performance following the first data.

\subsection{Full angular fit}
\label{sec:decay_model}
With the model above we can generate an ensemble of experiments corresponding
to a given integrated luminosity. In each of these experiments we can use a
general minimiser to find the spin value parametrisation that best corresponds
to the data. Each fit has in total 27 parameters; 26 from the signal described
above and a single parameter to describe the level of the simplified background
model. From the ensemble of experiments, estimates of the experimental
uncertainties can be made and any biases introduced can be studied. For each
dataset, the extracted spin amplitude components were used to calculate the
value of each angular observable as a function of \qsq. In total we created an
ensemble of 1000 experiments and will thus at a given value of \qsq get 1000
different determinations of a given observable. By looking at the point where
33\% and 47.5\% of results lie within either side of the median of the results
we can form asymmetric $1\sigma$ and $2\sigma$ errors. Connecting these at
different \qsq values gives us $1\sigma$ and $2\sigma$ bands for the
experimental errors on the observable. An illustration of the method in
\fig{fig:gamma_ex} shows the experimental sensitivity to the width
distribution relative to the normalisation point which was arbitrarily chosen
as 3.5\gevgev. The inner and outer bands correspond to 
1$\sigma$ and 2$\sigma$ experimental
errors with statistics corresponding to a 10\invfb dataset from \lhcb. The
dashed line is the theoretical input and the red line the central value of the
ensemble experiments. The difference between these two lines is caused by 
limitations imposed by the second order polynomial assumption. 
As it is is well inside the 1$\sigma$ band this is not problem.
\FIGURE[t]{\includegraphics[width=0.60\textwidth]{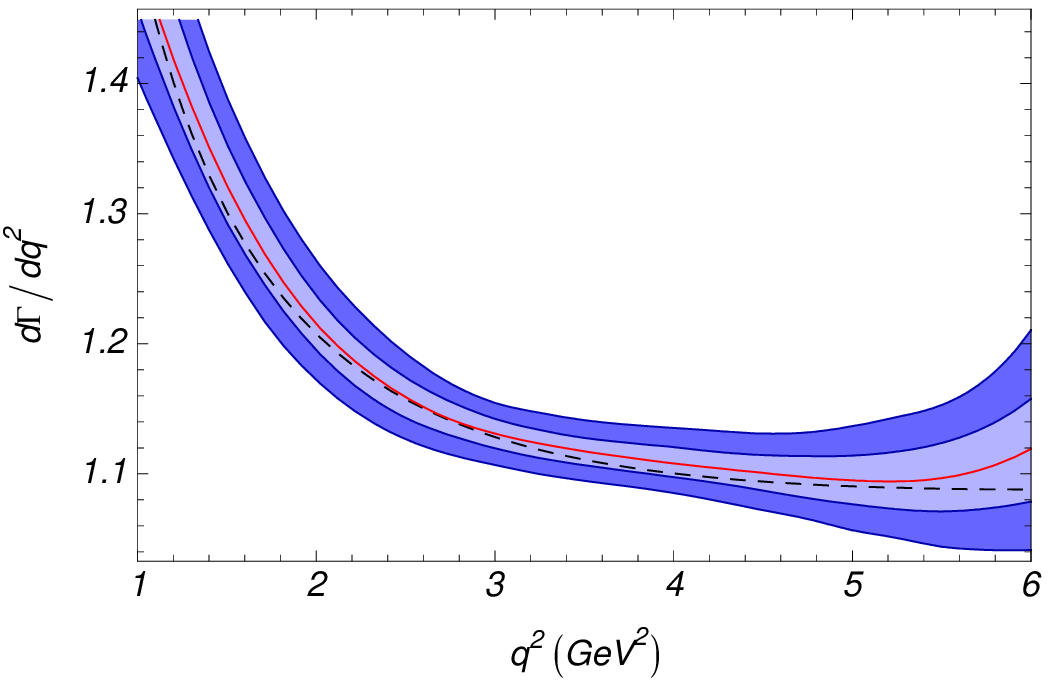}
  \caption{\label{fig:gamma_ex}The experimental sensitivity to
    $\frac{d\Gamma}{dq^2}$ with the SM as input. The inner and outer bands
    correspond to 1$\sigma$ and 2$\sigma$ experimental errors with statistics
    corresponding to a 10\invfb dataset from \lhcb. The black dashed line is the
    theoretical input and the red solid line the central value of the ensemble
    experiments.}
}

The experimental sensitivity to the observables introduced in 
Sec.~\ref{sec:observables} will be
presented in Sec.~\ref{sec:pheno} within the phenomenological analysis to
allow for an easy comparison of experimental and theoretical errors.

\subsection{Comparisons with fits to projections}
The full angular fit gives access to angular observables not accessible in
other ways. However, \AFB, \AT{2}, \FL and $\Aim$\footnote{\Aim is defined as
  $\Aim=\frac{\Imag{A_{\perp L} A_{\| L}^*} + \Imag{A_{\perp R} A_{\| R}^*}}
  {|A_{0}|^2 + |A_{\perp}|^2 +|A_{\|}|^2}$ and is included for completeness
  here. It is not of importance for the measurement of right handed currents.}
can be extracted from distributions in just a single angle after integration
over the other 2 in \eq{diff:four-fold}:
\begin{subequations}
  \label{eq:projections}
  \begin{eqnarray}
    \label{eq:dGammadPhi}
    \frac{d\Gamma^\prime}{d\phi} & = & \frac{\Gamma^\prime}{2\pi}\left(
      1 + \frac{1}{2} (1 - \FL) \AT{2} \cos 2\phi + 
      \Aim \sin 2\phi
    \right), \\
    \label{eq:dGammadThetal}
    \frac{d\Gamma^\prime}{d\thetaL} & = & \Gamma^\prime \left(
      \frac{3}{4}\FL \sin^2\thetaL + 
      \frac{3}{8} (1-\FL) (1+\cos^2\thetaL) +
      \AFB\cos\thetaL 
    \right)\sin\thetaL\, , \\
    \label{eq:dGammadThetaK}
    \frac{d\Gamma^\prime}{d\thetaK} & = &
    \frac{3\Gamma^\prime}{4} \sin\thetaK \left(
      2 \FL \cos^2\thetaK + (1-\FL) \sin^2\thetaK
    \right),
  \end{eqnarray}
\end{subequations}
where $\Gamma^\prime = b + 4c$. This method was investigated for \lhcb
in~\cite{Egede:2007zz}. The observables appear linearly in the expressions so
the fits can be performed on data binned in \qsq. The value extracted from
these fits is then a $\frac{d\Gamma}{dq^2}$ weighted average of each
parameter.

The full angular model described in \sec{sec:decay_model} was used to generate
data sets which were then fit simultaneously using the distributions in \eq{eq:projections}.
The treatment of background and the $m_{\B,K\pi}$ distributions were the same
as in the full angular model. For a direct comparison between this method and
the full angular fit, the \qsq dependent values of the observables were
averaged using a weighted mean,
\begin{equation}
  \label{eq:averageAT}
  \AT{i} = \frac{\int_{q^2_{\min}}^{q^2_{\max}}
                    \frac{d\Gamma}{d\qsq} \AT{i}(\qsq)}
                {\int_{q^2_{\min}}^{q^2_{\max}}
                    \frac{d\Gamma}{d\qsq}}\, .
\end{equation}
The central values produced for the full angular approach in this case show
some small biases due to the breakdown of the polynomial ansatz at the edges
of the \qsq distribution, however this is still well below the statistical
error expected with 10\invfb of data from \lhcb. The power of the full angular
fit is striking for \AT{2} where the resolution is above a factor 2 better
compared to fitting the projections. This can easily be understood in terms of
the $(1-\FL)$ factor in \eq{eq:dGammadPhi}, where \FL is large in the SM. 

For all the observables where a comparison can be made, we see that the full
angular fit provides improvements in the resolution of between 15\% and 60\%.

In the full angular fit we can calculate the position of the zero crossing for
the forward-backward asymmetry, \zerocrossing. We illustrate the distribution
of results obtained from the ensemble of datasets in \fig{fig:zerocrossing}
where a resolution, assuming the SM as input, of 0.17\gevgev is
obtained. Alternatively we can perform the simpler task of binning the data in
1\gevgev bins and then in each bin perform simultaneous fits to the three
angular projections. The value of \AFB is extracted by performing a straight
line fit in the range $2-6\gevgev$ to the \AFB values found in each \qsq bin.
This gives us, with exactly the same assumptions for how background and
acceptance are treated, a resolution of 0.24\gevgev. So also in this case we
see an improvement of 30\% in the statistical power by performing a full
angular fit.
\FIGURE{\includegraphics[width=0.60\textwidth]{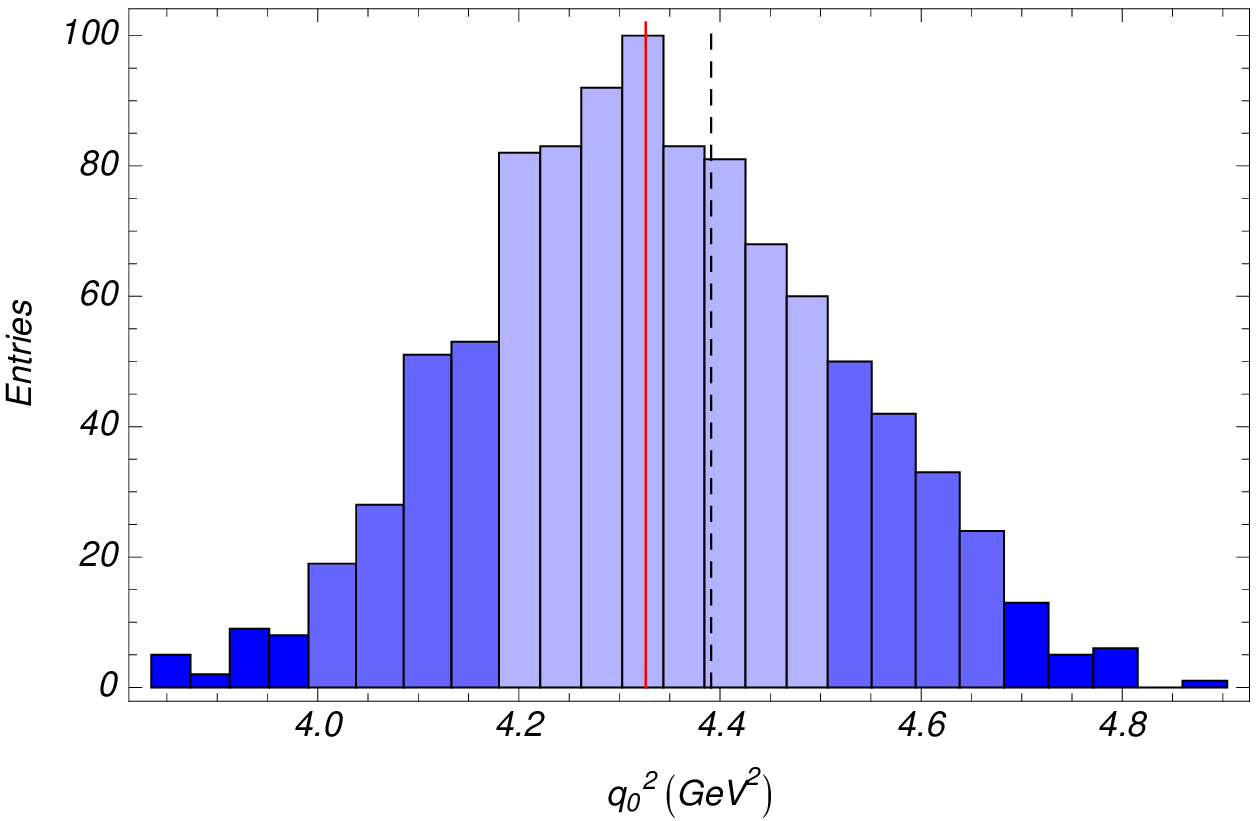}
  \caption{\label{fig:zerocrossing}The distribution in the
    determination of the zero crossing of \AFB from an ensemble of datasets
    created from the toy Monte Carlo model with statistics corresponding to
    10\invfb at \lhcb. The edge of the inner (light blue) and middle (medium
    blue) regions correspond to 1$\sigma$ and 2$\sigma$ experimental
    errors. The solid (red) line is the median of the fitted values and the
    dashed (black) line is the input value from the SM theory
    predictions. From the figure we see a resolution of 0.17\gevgev.}
}

The comparisons made here demonstrate that there is significant advantage in
performing the full angular fit once the data sets are large enough. Using the
simplified model described here it is possible to use this approach even with
a smaller 2\invfb data set. In reality, detector effects not accounted for
such as angular acceptance will complicate the process and a proper full
angular analysis may not be possible with data sets this small. However, we
have shown that with the signal and background statistics at \lhcb a full
angular analysis is possible once the detector effects have been properly
understood.

\section{Phenomenological analysis}
\label{sec:pheno}

In this section we present our phenomenological analysis of the old and new
observables in the SM and in extensions of the SM with new right-handed
currents.  The latter can be done in a model independent way by introducing
the chiral partners of the SM Wilson coefficients \Ceff{7}, \Ceff{9}, and
\C{10}\footnote{We note here that the impact of \Ceff{9} and $\C{10}$ and
  their chiral partners is rather small compared with \Cpeff{7} in the
  low-$q^2$ region, due to the 2 ${\hat m_b}/{\hat s}$ factor in the matrix
  element and the experimental constraints from the inclusive decay $B\to X_s
  \ell^+\ell^-$.}.  This general new-physics scenario can be realized for
example via gluino-mediated FCNC in the general R-parity conserving MSSM.

\subsection{Preliminaries}
Our analysis is based on the numerical input as summarized in
Table~\ref{tabInput}. Regarding the form factor value we follow
Ref.~\cite{Beneke:2004dp} and use the value fixed by experimental data.
%%%%%%%%%%%%%%%%%%%%%%%%%%%%%%%%%%%%%%%%%%%%%%%%%%%%%%%%%%%%%%%%%%%
\TABLE{\centerline{\parbox{14cm}{\caption{\label{tabInput}
Summary of input parameters and estimated uncertainties.}}}
     \begin{tabular}{| l l| l l |} 

\hline 
\hline  
\rule[-2mm]{0mm}{7mm}
     $\!\!m_B$                           &$5.27950 \pm 0.00033\,{\rm GeV}$ & 
     	$\lambda$                       & $0.2262 \pm 0.0014$ \\
     
     $m_K$                           &$0.896 \pm 0.040\,{\rm GeV}$ & 
         $A$                             & $0.815 \pm 0.013$ \\
	
     $M_W$                           & $80.403\pm 0.029\,{\rm GeV}$  & 
        $\bar\rho$                      & $0.235 \pm 0.031$ \\
        
     $M_Z$				&$91.1876 \pm 0.0021\,{\rm GeV}$ &
       $\bar\eta$                      & $0.349 \pm 0.020$
        
   \\[0.15cm]
\hline
\rule[-2mm]{0mm}{7mm}
     $\!\!\hat m_t(\hat m_t)$            & $167 \pm 5$~GeV &
      $\Lambda_{\rm QCD}^{(n_f=5)}$   & $220 \pm 40\,{\rm MeV}$\\
    
     $m_{b,\rm PS}(2\,\mbox{GeV})$   & $4.6 \pm 0.1$~GeV &
      $\alpha_{s}(M_Z)$ & $0.1176\pm 0.0002$ \\

     $m_c$                           & $1.5 \pm 0.2$~GeV & 
      $\alpha_{\rm em}$               & $ 1/137$

\\[0.15cm]
\hline
\rule[-2mm]{0mm}{7mm}
     $\!\!f_B$                           & $200 \pm 30$~MeV &
      $a_1(K^*)_{\perp,\,\parallel}$    &  $0.10 \pm 0.07$\\
   
     $f_{K^*,\perp}$  & $175 \pm 25$~MeV &
      $a_2(K^*)_{\perp}$    &  $0.13 \pm 0.08$\\
        
     $f_{K^*,\parallel}$             & $217 \pm 5$~MeV &
      $a_2(K^*)_{\parallel}$    & $0.09 \pm 0.05$

\\[0.15cm]
\hline
\rule[-2mm]{0mm}{7mm}
     $\!\!m_B\,\xi_{K^*,\parallel}(0)/(2m_{K^*})$   & $ 0.47 \pm 0.09$ &
       	$\lambda_{B,+} (1.5{\rm GeV})$  & $0.485 \pm 0.115 \,{\rm GeV}$\\
     $\xi_{K^*,\perp}(0)$               & $  0.26 \pm 0.02$ &&
\\[0.15cm]
\hline
\hline  
       \end{tabular} 
}
%%%%%%%%%%%%%%%%%%%%%%%%%%%%%%%%%%%%%%%%%%%%%%%%%%%%%%%%%%%%%%%%%%%
Moreover, 
we introduce four representative benchmark points  of supersymmetry
with non-minimal flavour violation in the down squark sector which were 
already used in Ref.~\cite{Lunghi:2006hc}. The most important flavour diagonal
parameters are fixed as follows: $\tan\beta =5$, $\mu=M_1=M_2=M_{H^+}= m_{\tilde u_R}=1\tev$. 
Note that we choose a low value for $\tan\beta$; this shows that we 
do not need to rely on a
large-$\tan \beta$ to see an effect, and ensures automatic fulfilment of the
constraint coming from $B_s \to \mu^+ \mu^-$.
Furthermore, we make the assumption that all the entries in $m^2_{u,LR}$ and $m^2_{d,LR}$
vanish, with the exception of the one that corresponds to
$\left(\delta_{LR}^{d}\right)_{32}$. The remaining parameters of the four benchmark points 
correspond to two different scenarios and are fixed as follows\footnote{We follow here the conventions of Ref.~\cite{Foster:2005wb}}: 
\begin{itemize}
\item Scenario A: $m_{\tilde g} = 1\tev$ and $m_{\tilde d} \in
  [200,1000]\gev$. The only non-zero mass insertion is varied between
  $-0.1\leq \left(\delta_{LR}^{d}\right)_{32} \leq 0.1$. For all parameter
  sets the compatibility with other $B$ physics constraints, the electroweak
  constraints, constraints from particle searches, and also with the vacuum
  stability bounds is verified~\cite{Lunghi:2006hc}.  The curves denoted by
  (a) and (b) correspond respectively to $m_{\tilde g}/m_{\tilde d}=2.5$,
  $\left(\delta_{LR}^{d}\right)_{32}=0.016$ and $m_{\tilde g}/m_{\tilde d}=4$,
  $\left(\delta_{LR}^{d}\right)_{32}=0.036$. We will refer to this case as the
  large-gluino and positive mass insertion scenario. In terms of the effective
  Wilson coefficients at $m_b$, model (a) corresponds to $(\Ceff{7}, \Cpeff{7})=(-0.32,
  0.16)$ and (b) to $(-0.32, 0.24)$. This should be compared to the SM value
  of $(\Ceff{7}, \Cpeff{7})=(-0.31, 0.00)$.
\item Scenario B: $m_{\tilde d} = 1\tev$ and $m_{\tilde g} \in
  [200,800]\gev$. The mass insertion is varied in the same range as Scenario
  A. The curves denoted by (c) and (d) correspond respectively to $m_{\tilde
    g}/m_{\tilde d}=0.7$, $\left(\delta_{LR}^{d}\right)_{32}=-0.004$ and
  $m_{\tilde g}/m_{\tilde d}=0.6$,
  $\left(\delta_{LR}^{d}\right)_{32}=-0.006$. We will refer to this case as
  the low-gluino mass (although large squark mass would be more appropriate)
  and negative mass insertion scenario. In this case the corresponding
  effective Wilson coefficients are $(\Ceff{7}, \Cpeff{7})=(-0.32, -0.08)$ for
  (c) and $(-0.32, -0.13)$ for (d).
\end{itemize}

Notice that we have changed curve (c) with respect to Ref.~\cite{Lunghi:2006hc} reducing its
corresponding mass insertion to avoid any conflict with vacuum stability or
colour breaking constraints \cite{casas}.

Finally, we emphasize again that the validity of our theoretical predictions
is restricted to the kinematic region in which the energy of the $K^*$ is of
the order of the heavy quark mass. So we restrict our analysis to the
low-$q^2$ region from 1\gevgev to 6\gevgev. In the region below 1\gevgev the
QCDf/SCET results are questioned by the presence of very light resonances.

\subsection{Results}
We present our results on the observables \AT{2}, \AT{3}, \AT{4}, \AFB and
$F_L$ in the Figs.~\ref{fig:AT2}--\ref{fig:FL} (for definitions see
Sec.~\ref{sec:clean}). For all the observables we plot the theoretical
sensitivity on the left hand side of each Figure.
\begin{itemize}
\item 
  The thin dark line is the central NLO result for the SM and the narrow inner
  dark (orange) band that surrounds it corresponds to the NLO SM uncertainties
  due to both input parameters and perturbative scale dependence. Light grey
  (green) bands are the estimated $\Lambda/m_b \pm 5\%$ corrections for each
  spin amplitude (as given in Eq.~\ref{lambdamb}) while darker grey (green)
  ones are the more conservative $\Lambda/m_b \pm 10\%$ corrections. The
  curves labelled (a)--(d) correspond to the four different benchmark points
  in the MSSM introduced above.
\item The experimental sensitivity for a dataset corresponding to 10\invfb of
  \lhcb data is given in each figure on the right hand side.  Here the solid
  (red) line shows the median extracted from the fit to the ensemble of data
  and the dashed (black) line shows the theoretical input distribution.  The
  inner and outer bands correspond to 1$\sigma$ and 2$\sigma$ experimental
  errors.
\end{itemize}
\FIGURE{\includegraphics[width=0.49\textwidth]{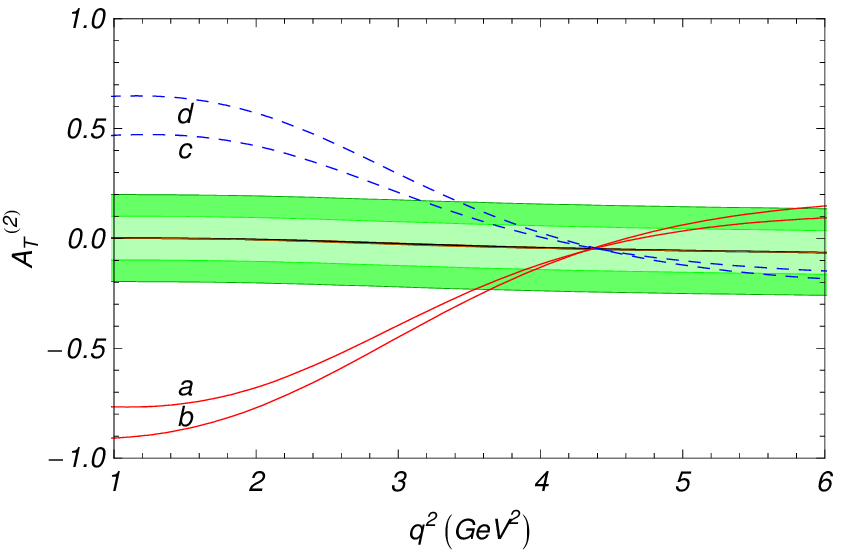}
  \includegraphics[width=0.49\textwidth]{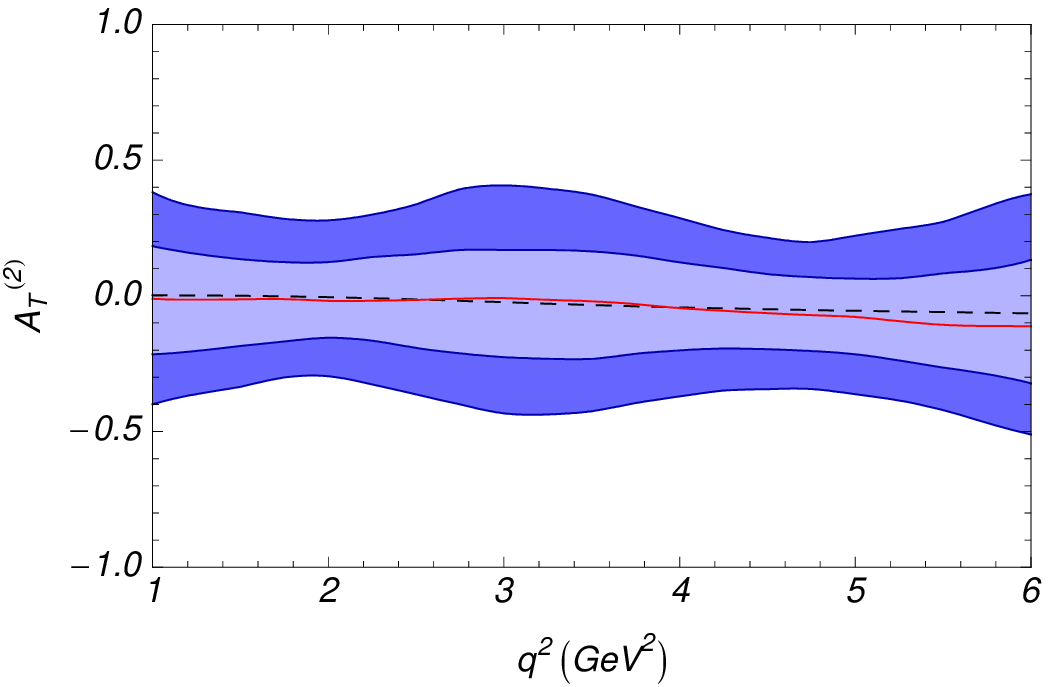}
  \caption{For \AT{2} we compare the theoretical errors (left) with the
    experimental errors (right) as a function of the squared dimuon mass. For
    the theory, the narrow inner dark (orange) bands correspond to the NLO
    result for the SM including all uncertainties (except for $\Lambda/m_b$)
    as explained in the text. Light grey (green) bands include the estimated
    $\Lambda/m_b$ uncertainty at a $\pm 5\%$ level and the external dark grey
    (green) bands correspond to a $\pm 10\%$ correction for each spin
    amplitude. The curves labelled (a)--(d) correspond to different SUSY
    scenarios as explained in the text. For the experimental aspects the inner
    and outer bands correspond to 1$\sigma$ and 2$\sigma$ statistical errors
    with a yield corresponding to a 10\invfb dataset from \lhcb.}
\label{fig:AT2}
}
\FIGURE{\includegraphics[width=0.49\textwidth]{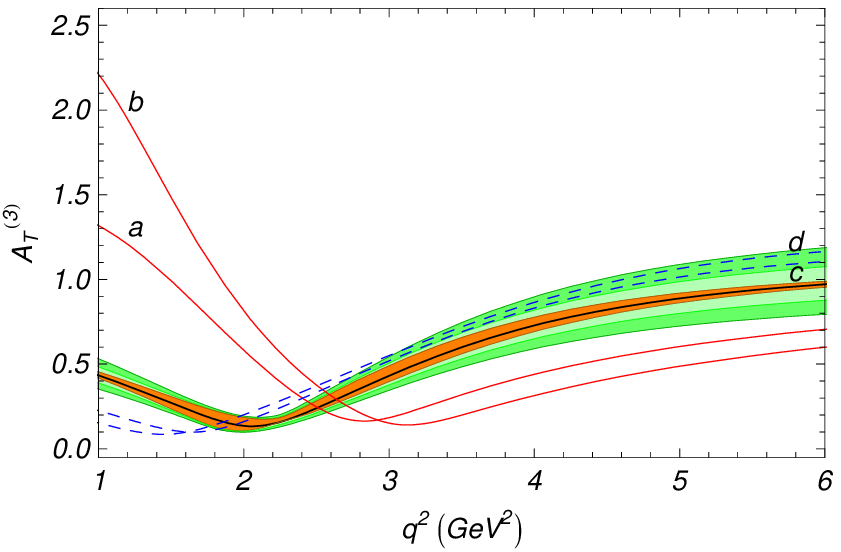}
  \includegraphics[width=0.49\textwidth]{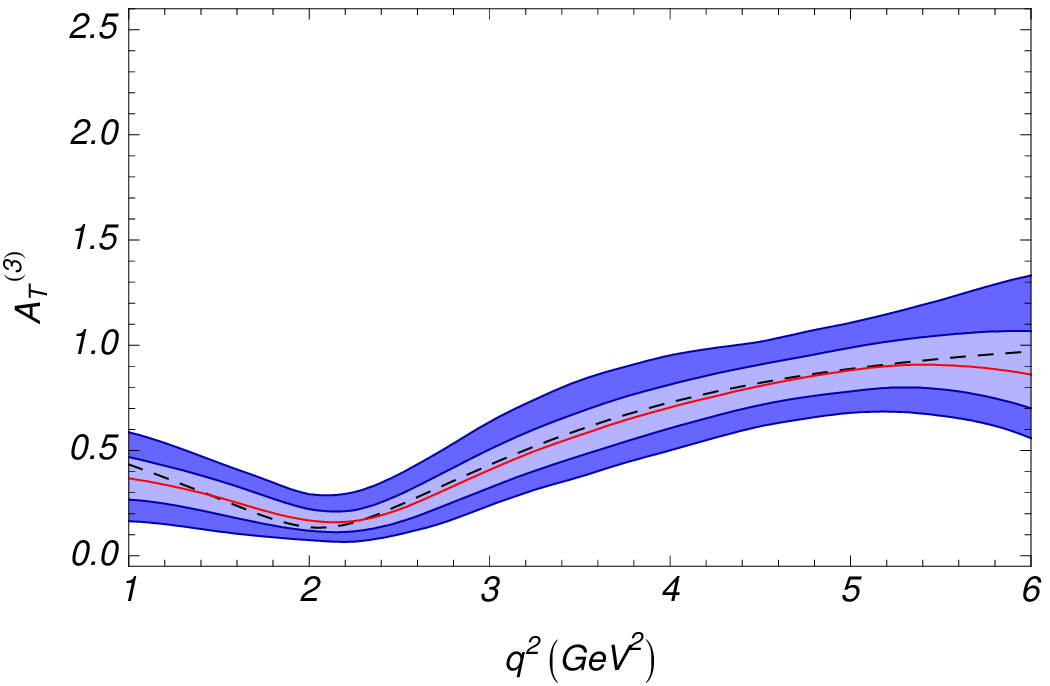}
  \caption{For the new observable \AT{3} we compare the theoretical errors
    (left) with the experimental errors (right). See the caption of
    Fig.~\protect\ref{fig:AT2} for details.}
\label{fig:AT3}
}
\FIGURE{\includegraphics[width=0.49\textwidth]{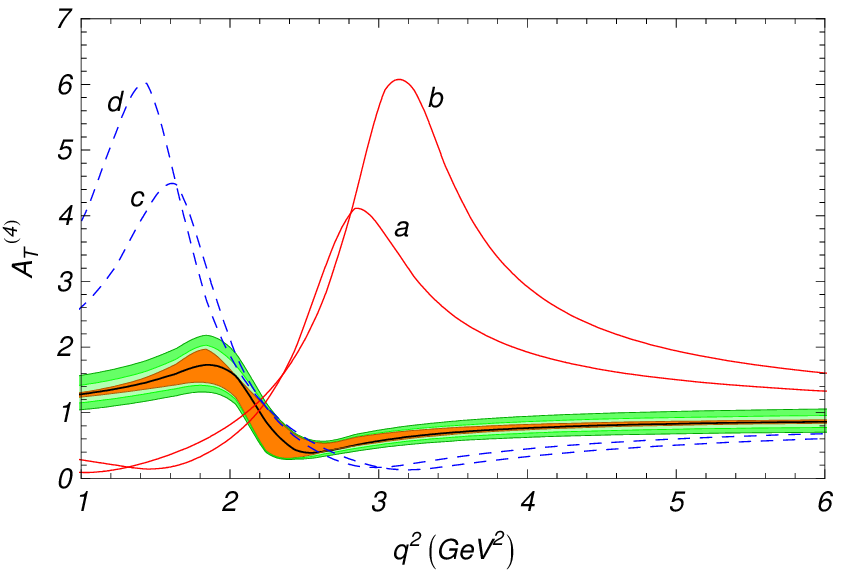}
  \includegraphics[width=0.49\textwidth]{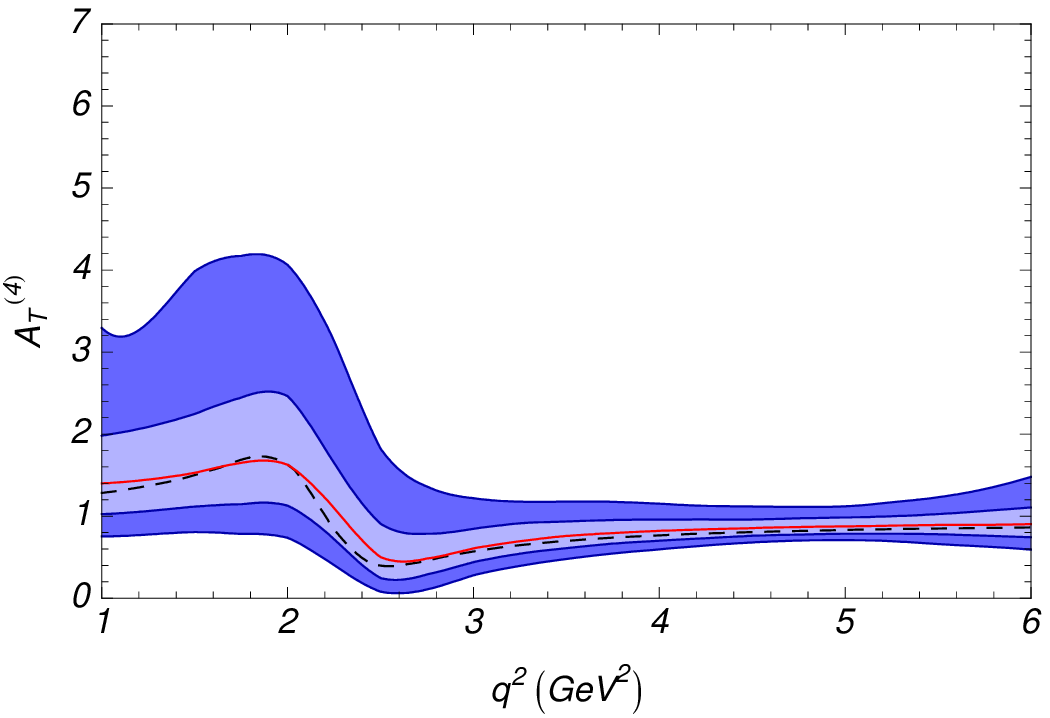}
  \caption{For the new observable \AT{4} we compare the theoretical
    errors (left) with the experimental errors (right). See the caption of
    Fig.~\protect\ref{fig:AT2} for details.}
  \label{fig:AT4}
}
\FIGURE{\includegraphics[width=0.49\textwidth]{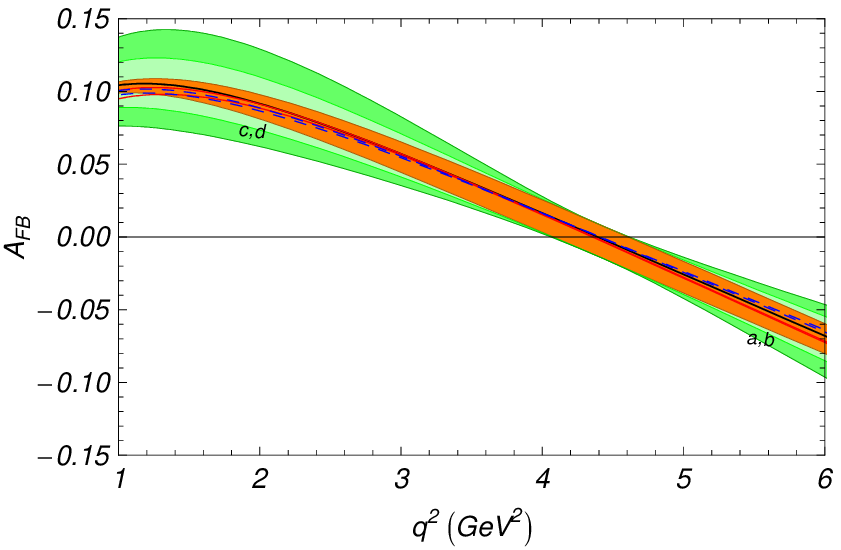}
  \includegraphics[width=0.49\textwidth]{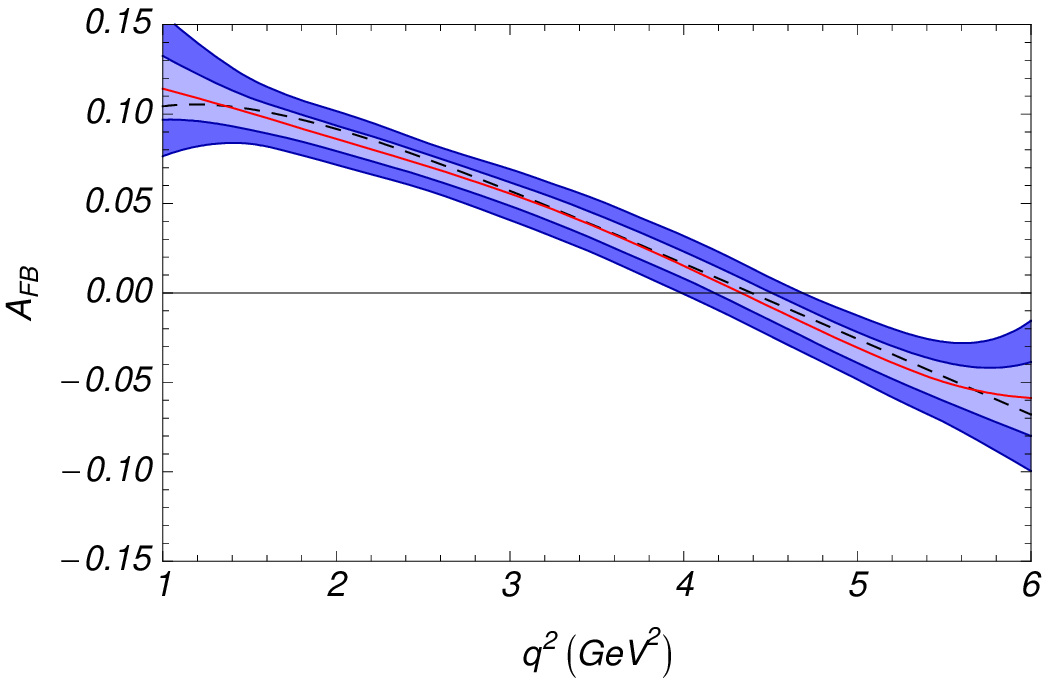}
  \caption{For $A_{FB}$ we compare the theoretical errors (left) with the
    experimental errors (right). See the caption of Fig.~\protect\ref{fig:AT2}
    for details.}
  \label{fig:Afb}
}
\FIGURE{\includegraphics[width=0.49\textwidth]{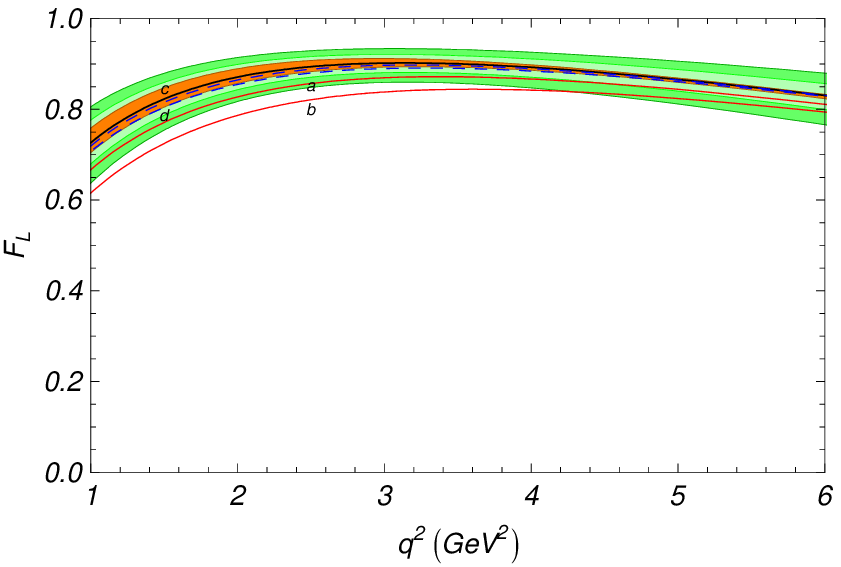}
  \includegraphics[width=0.49\textwidth]{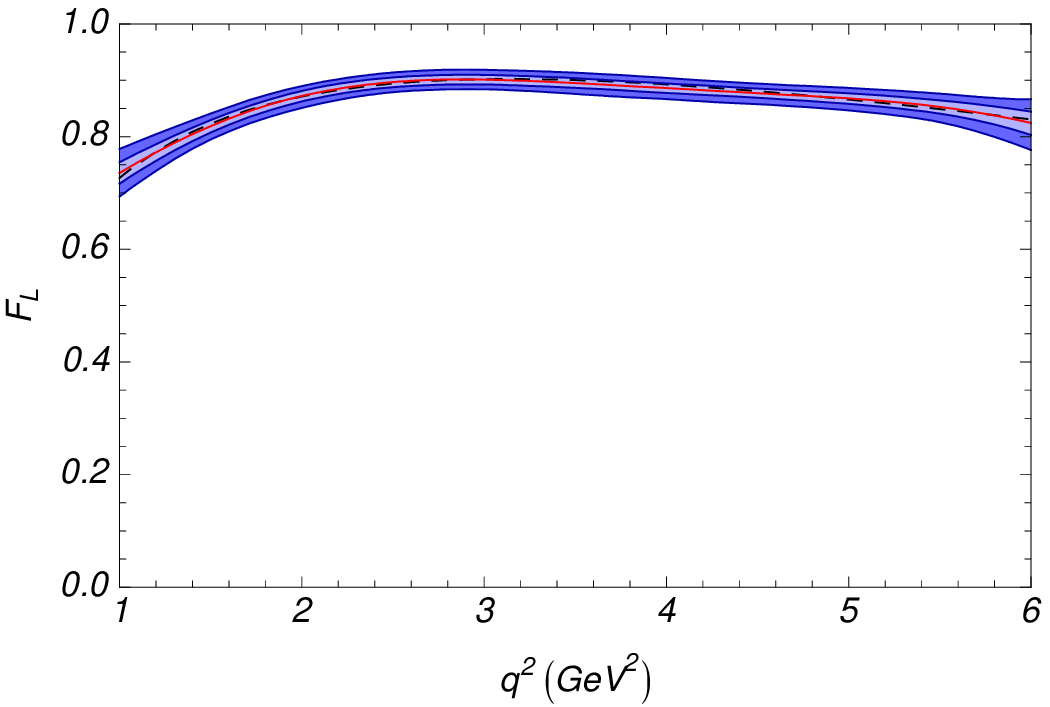}
  \caption{For $F_L$ we compare the theoretical errors (left) with the
    experimental errors (right). See the caption of Fig.~\protect\ref{fig:AT2}
    for details.}
  \label{fig:FL}
}

Let us start with some concrete observations on the new observables 
\AT{3} and \AT{4}. They  offer sensitivity to the longitudinal
spin amplitude $A_{0 L,R}$ in a controlled way compared to the old observables 
$F_L$ and $\alpha_K^*$: the dependence on both the parallel and perpendicular 
soft form factors $\xi_\|(0)$ and $\xi_\perp(0)$ cancels at LO. A residual of this
dependence may appear at NLO, but as shown in Figs.~\ref{fig:AT3} and \ref{fig:AT4},  
it is basically negligible. 
It is also remarkable that for \AT{3} and \AT{4} at low \qsq the
impact of this uncertainty is less important than the 
uncertainties due to input parameters and scale dependence. 

The peaking structure in \AT{4} as a function of \qsq for the benchmark MSSM
points is due to the different way \Cpeff{7} enters numerator and denominator;
the numerator has a positive slope in the region of the peak, while the
denominator has a minimum at the same point. If one uses the simplified L0 expressions from Eqs.~\ref{LEL:tranversity:perp}--\ref{LEL:tranversity:zero}
the denominator is exactly zero, generating
an infinity at the point of the peak; however, once NLO QCDf is included
the zero in the denominator is lifted and the result is a curve with a peak instead.

The new observables \AT{3} and \AT{4} also present a different sensitivity to
\Cpeff{7} via their dependence on  $A_{0 L,R}$ compared with \AT{2}.  
This may allow for a particularly interesting cross check of the sensitivity 
to this chirality flipped operator \Opep{7}; for instance, new contributions 
coming from tensor scalars and
pseudo-scalars will behave differently among the set of observables.

Another remarkable point that comes clear when comparing the set of clean
observables \AT{2}, \AT{3} and \AT{4} versus the old observables like $F_L$
concerns the potential discovery of NP, in particular of new right-handed currents.  The new observables share
the nice feature of \AT{2} that there are large deviations from the SM curve
from the ones of the four supersymmetric benchmark points. In case of \AT{2}
this is caused by the balance between the competing contributions of order
$1/\qsq$ and $1/q^4$ originating from the photon pole in the numerator and
denominator of \AT{2}, providing a strong sensitivity to \Cpeff{7}. This
sensitivity is near maximal around the 1\gevgev region precisely inside the
theoretically well controlled area. A large deviation from the SM for \AT{2}, \AT{3} or \AT{4} can thus show the presence of right-handed currents in a way that is not
possible with $F_L$ or $A_{FB}$.  In the latter cases the deviations from the
SM prediction of the same four representative curves are marginal.

In the experimental plots we find a good agreement between the central values
extracted from the fits and the theoretical input. Any deviations seen are
small compared to the statistical uncertainties, however the weakness of the
polynomial parametrisation, particularly at the extremes of the \qsq range,
can be seen. For much larger data sets this could be addressed by increasing
the order of the polynomials used. The experimental resolution for $F_L$ is
very good but with the small deviations from the SM expected this is not
helpful in the discovery of new right-handed currents. 
Comparing the theoretical and experimental figures for the other 
observables it can be seen that in
particular \AT{3} show great promise to distinguish between NP
models.

To further explore the power of the observables we can imagine that nature
corresponds to SUSY scenario (b). We create an ensemble of datasets from the
toy Monte Carlo model assuming model (b) as input and compare the results
to the SM prediction including the theoretical errors to get a feeling for how
significantly different from the SM prediction the results are. The results of
this are presented in
Figs.~\ref{fig:AT2andAT3compare}-\ref{fig:AFBandFLcompare}: It can be seen
that \AT{2}, \AT{3} and \AT{4} all show a remarkable separation between the
experimental error band and the SM prediction thus providing high sensitivity
to NP. For the SUSY scenario (b) chosen here, the deviation for \AFB and $F_L$
on the other hand is minor.
\FIGURE{\includegraphics[width=0.49\textwidth]{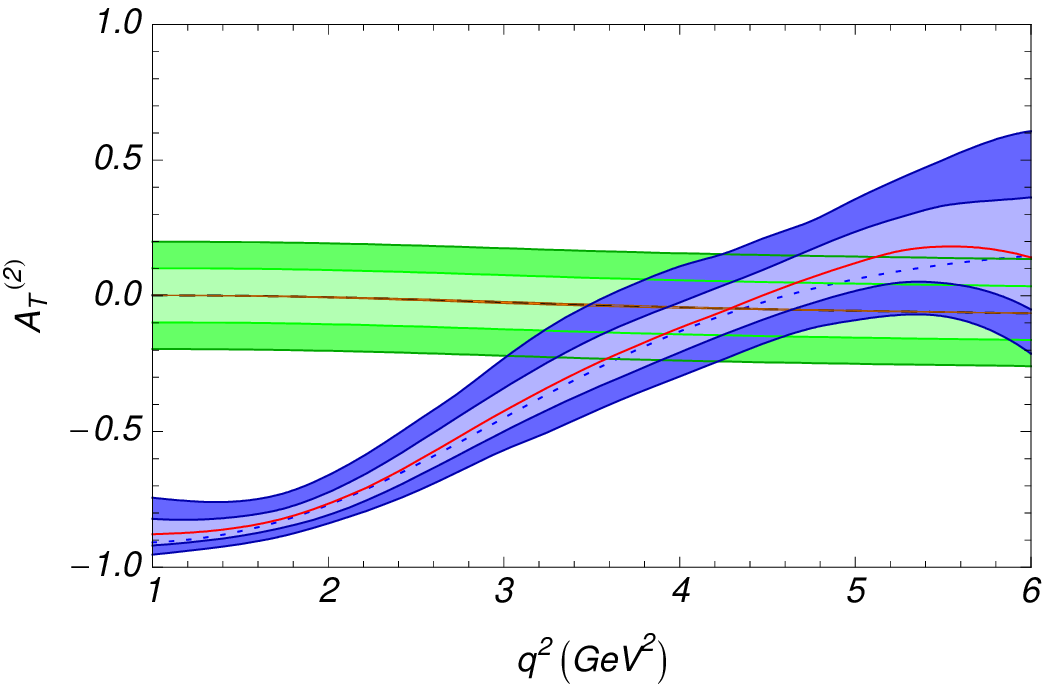}
  \includegraphics[width=0.49\textwidth]{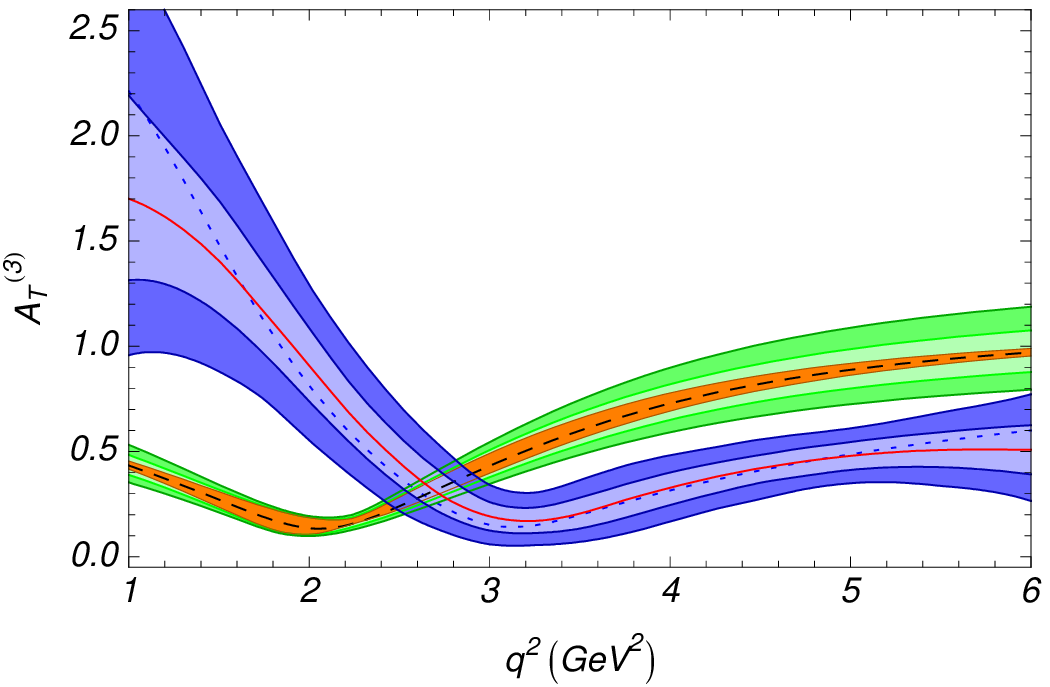}
  \caption{The experimental errors (blue, on top) assuming SUSY scenario (b)
    with large-gluino mass and positive mass insertion, is compared to the
    theoretical errors (green, below) assuming the SM. To the left for \AT{2}
    and the right for \AT{3}.}
  \label{fig:AT2andAT3compare}
}

\FIGURE{\hspace{1cm}\includegraphics[width=0.49\textwidth]{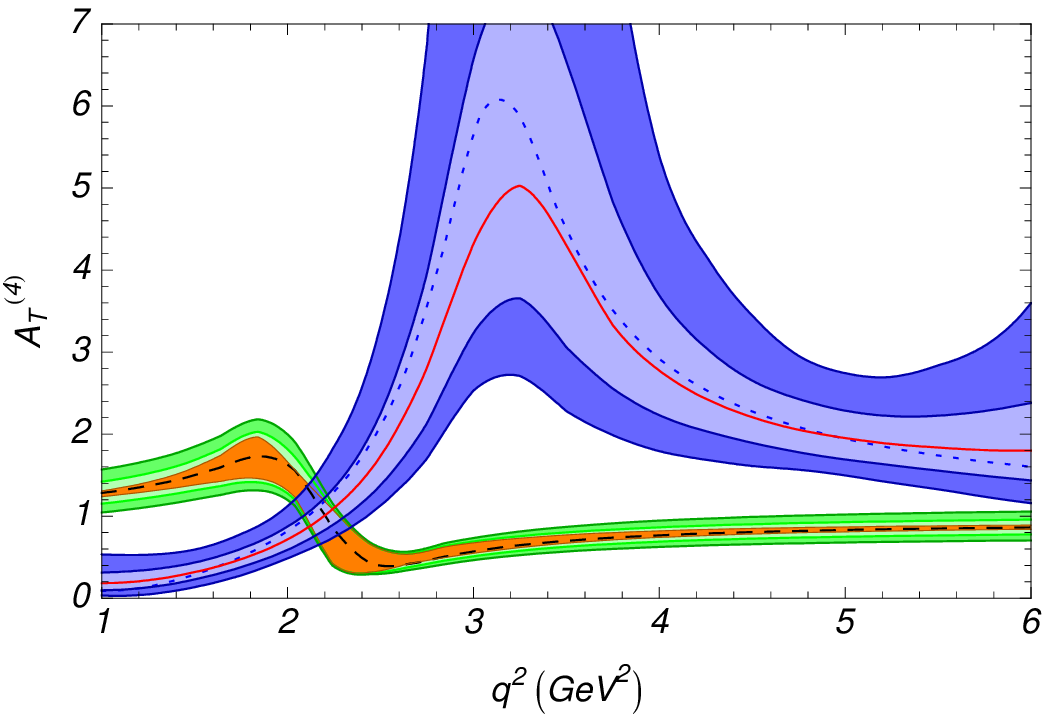}\hspace{1cm}
  \caption{The experimental errors (blue, on top) assuming SUSY scenario (b)
    with large-gluino mass and positive mass insertion, is compared to the
    theoretical errors (green, below) assuming the SM. Here the observable
    \AT{4} is considered.}
  \label{fig:AT4compare}
}

\FIGURE{\includegraphics[width=0.49\textwidth]{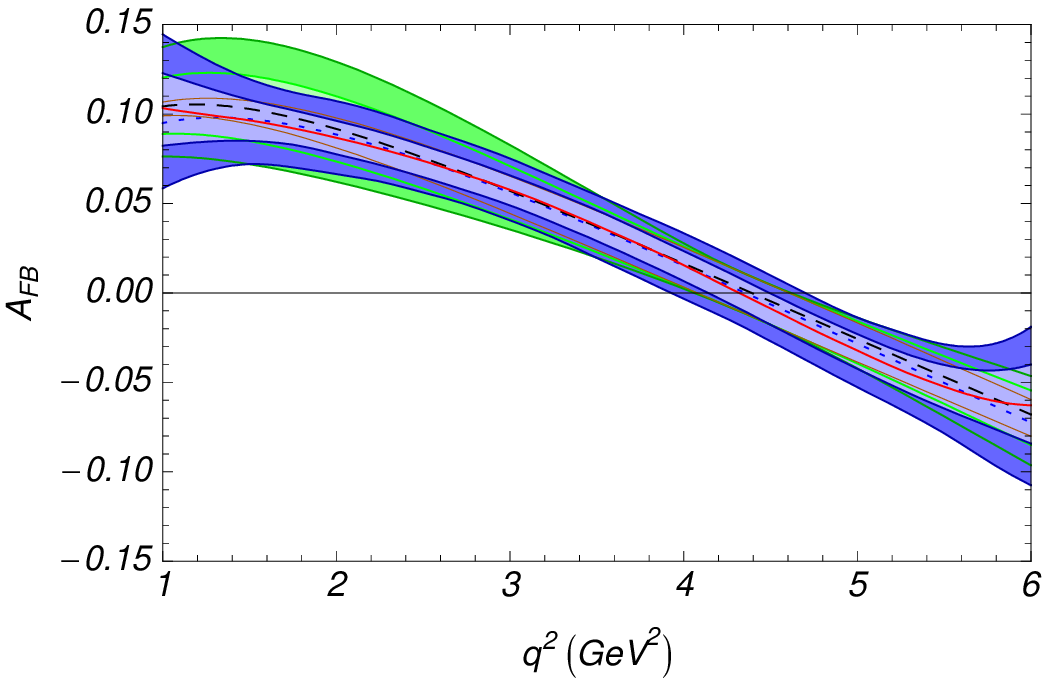}
  \includegraphics[width=0.49\textwidth]{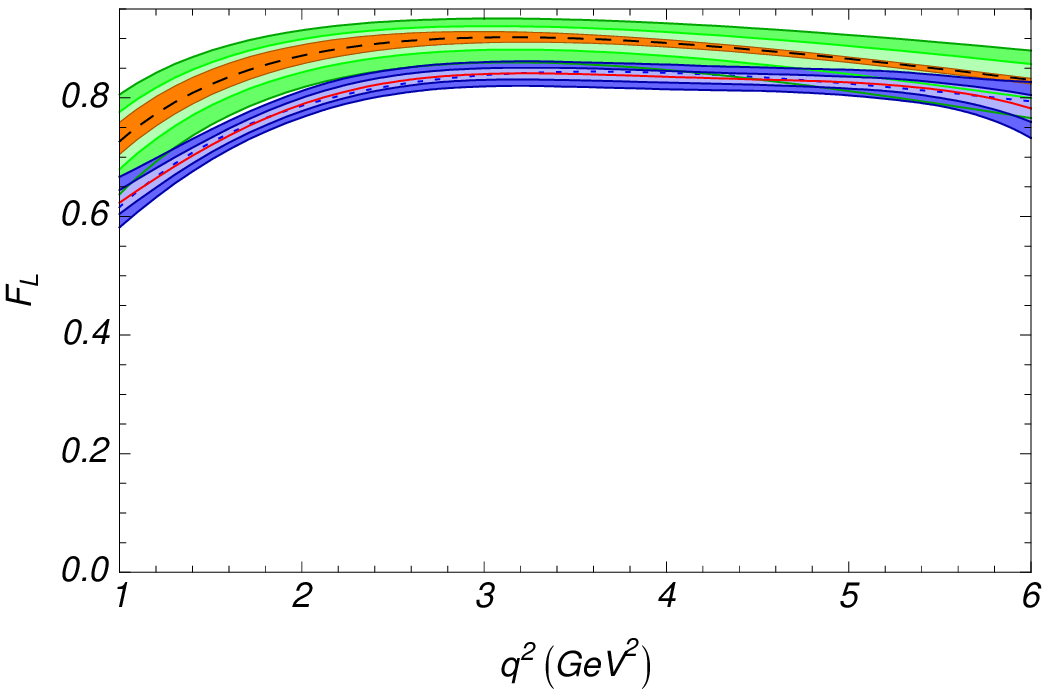}
  \caption{The experimental errors (blue, on top) assuming SUSY scenario (b)
    with large-gluino mass and positive mass insertion, is compared to the
    theoretical errors (green, below) assuming the SM. To the left for \AFB
    and the right for $F_L$.}
  \label{fig:AFBandFLcompare}
}

\FIGURE{\hspace{1cm}\includegraphics[width=0.49\textwidth]{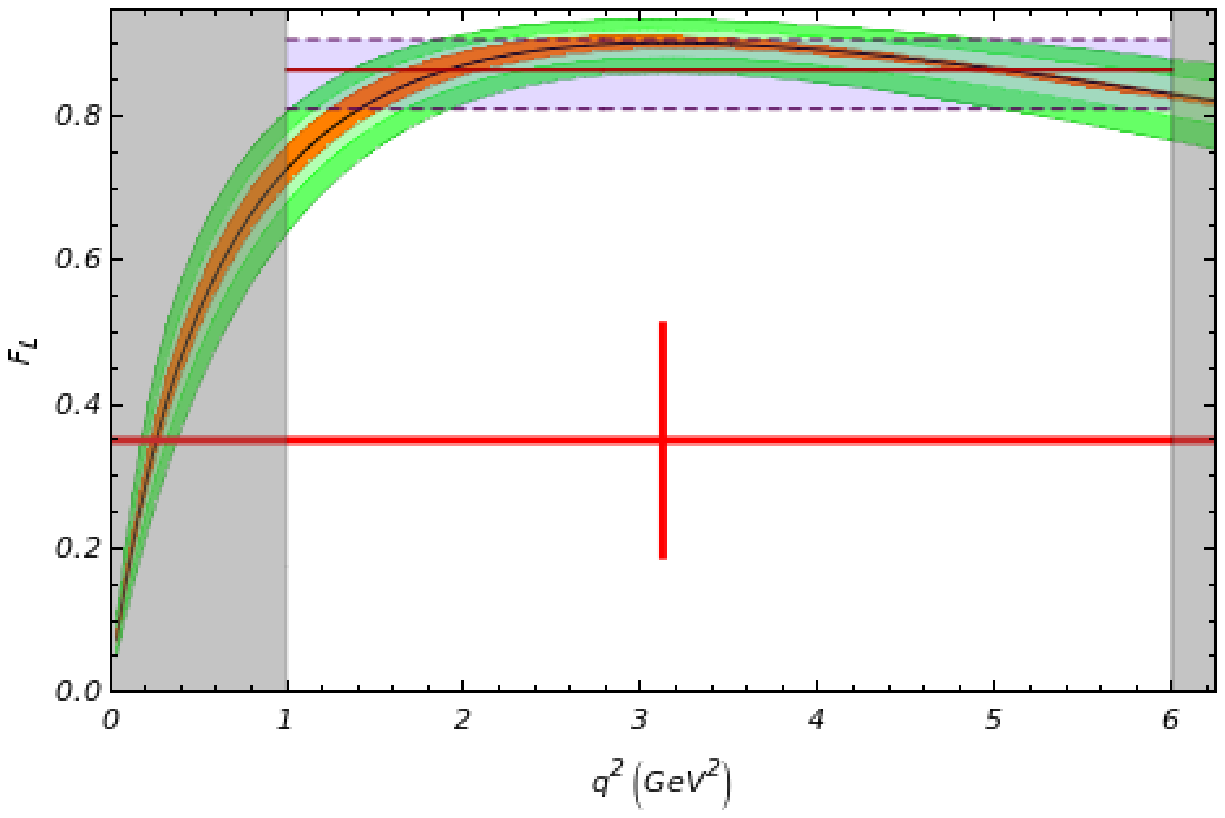}\hspace{1cm}
  \caption{Weighted SM average over the bin $q^2 \in [1\gevgev, 6\gevgev]$ and
    recent \babar\ measurement using the extended bin $q^2 \in [4 m_\mu^2,
    6.25\gevgev]$ (shown in grey).}
  \label{fig:FLbabar}
}
As mentioned in the introduction, the $B$ factories can already access some of
the angular observables using the projection-fit method described in
Sec.~\ref{sec:sensitivity}.  For example, recently the \babar\ collaboration
announced the first measurement of the longitudinal polarisation in the low
\qsq region as an average over the bin $q^2 \in [4m_\mu^2,
6.25\gevgev]$~\cite{Aubert:2008ju} (see Fig.~\ref{fig:FLbabar}):
\begin{equation}
F_L (q^2 \in [4m_\mu^2, 6.25\gevgev]) = 0.35 \pm 0.16_{\rm stat} \pm 0.04_{\rm syst}\, . 
\end{equation}
However, as mentioned before, the spectrum below 1\gevgev is 
theoretically problematic; moreover the rate and also the polarisation $F_L$ 
are changing dramatically around 1\gevgev. 
Therefore, we strongly recommend to 
use the standard bin from 1\gevgev to 6\gevgev. 
For future comparison we give here the theoretical  average, weighted over the rate, 
using  the bin, $q^2 \in [1\gevgev, 6\gevgev]$, based on our results: 
\begin{equation}
F_L (q^2 \in [1\gevgev, 6\gevgev]) = 0.86\pm0.05\, . 
\end{equation}
and refer to Fig.~\ref{fig:FL} for the future experimental sensitivity of the
\lhcb experiment. In Fig.~\ref{fig:FLbabar} we see the theoretical \qsq
distribution of \FL with the rate average overlaid.

\FIGURE{\includegraphics[width=0.49\textwidth]{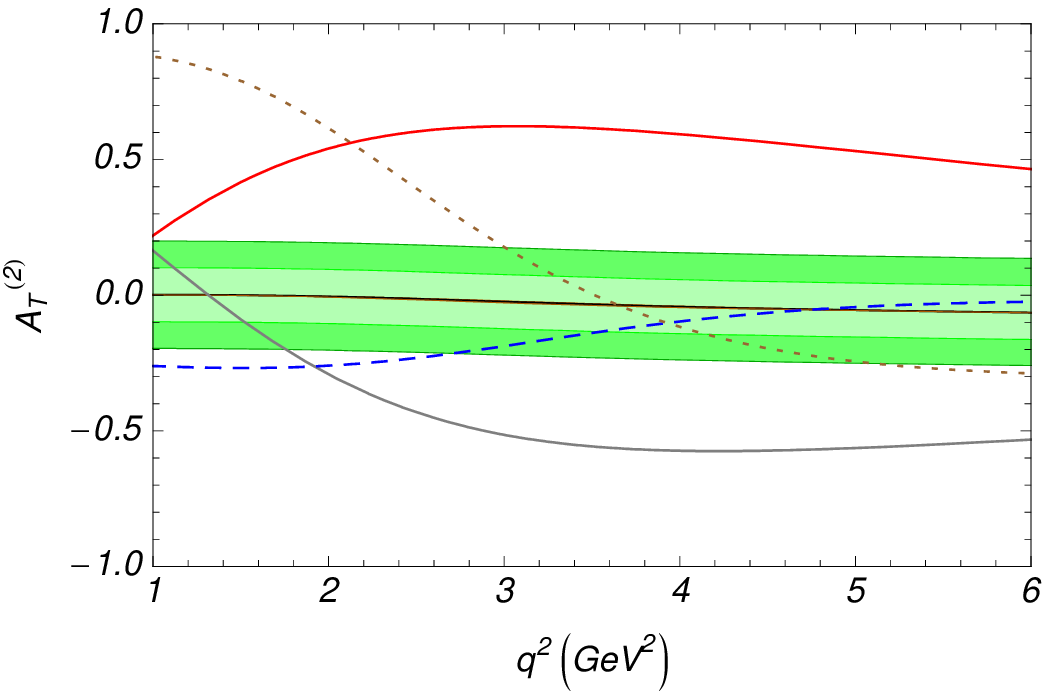}
\includegraphics[width=0.49\textwidth]{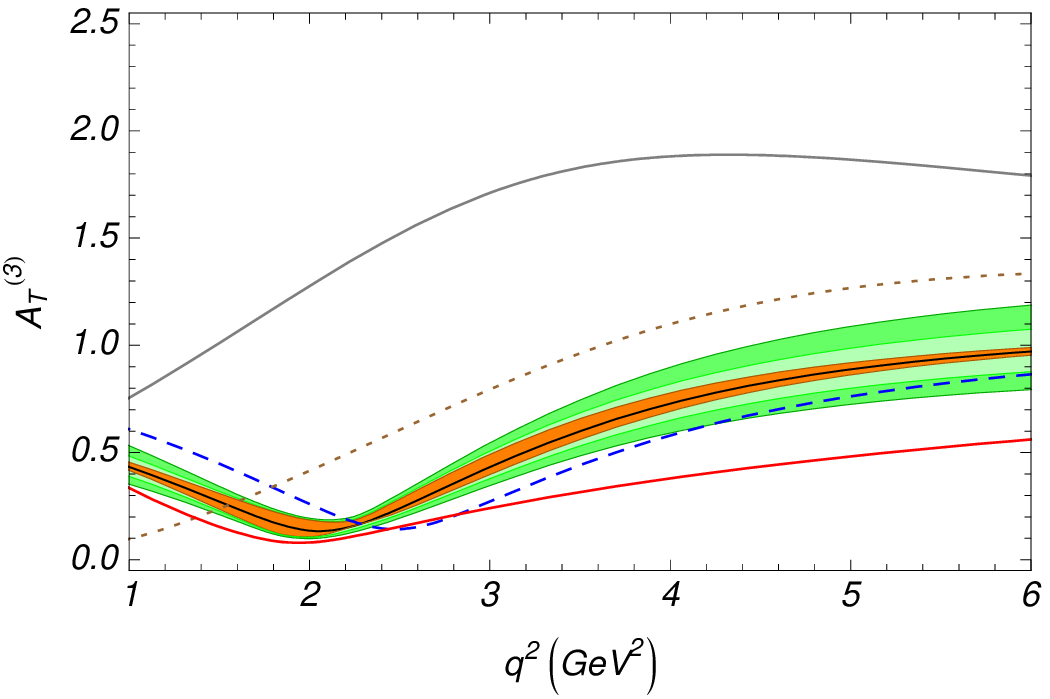}
\includegraphics[width=0.49\textwidth]{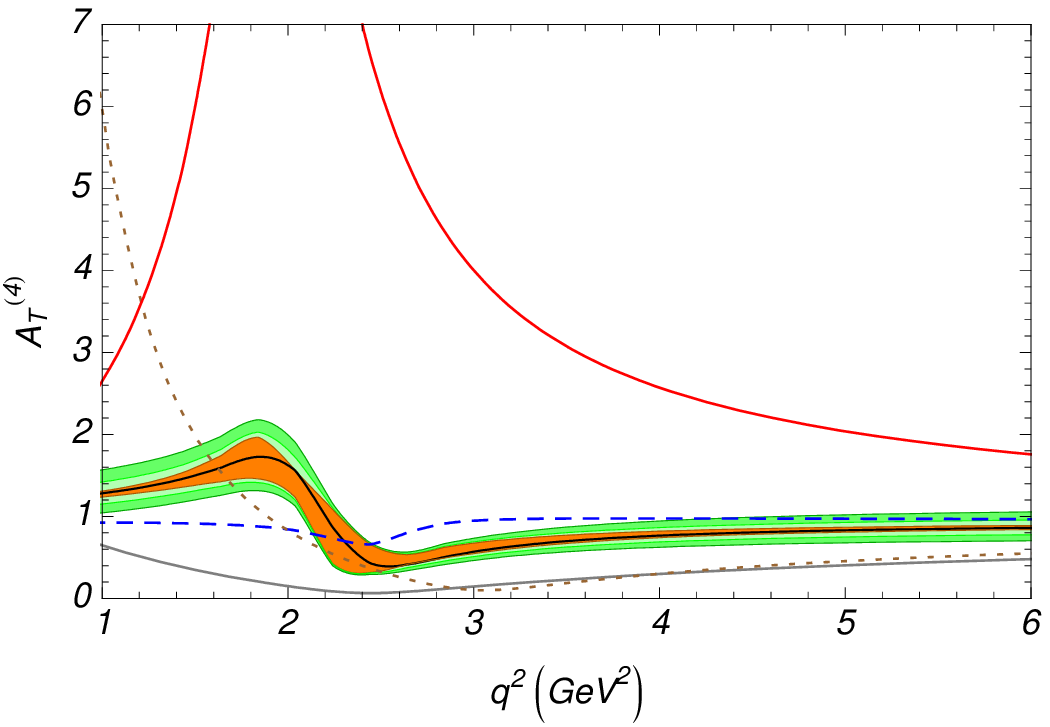}
\caption{The distribution of \AT{2}, \AT{3} and \AT{4} for four allowed
  combinations of \Ceff{7} and \Cpeff{7} following the model independent
  analysis of~\cite{Bobeth:2008ij}. The bands correspond to the SM and the
  theoretical uncertainty as described in Fig~\ref{fig:AT2}. The solid heavy
  (red) line corresponds to $(\Ceff{7}, \Cpeff{7})=(0.04,0.31)$, the solid
  light (grey) line $(-0.03,-0.32)$, the dashed (blue) line $(-0.35,0.05)$,
  and the dotted (brown) line $(-0.24,-0.19)$. Combining measurements in all
  three asymmetries will provide clear distinction between the different
  allowed regions.}
  \label{fig:ModelIndependent}
}
Rather than using the benchmark supersymmetry points for the illustration of
the power of the observables, one can also look at it from a model independent
point of view. For this we have taken four illustrative points from Fig.~2
in~\cite{Bobeth:2008ij} which are all allowed given the constraints from
present measurements of $b \to s$ transitions. In
Fig.~\ref{fig:ModelIndependent} the effect can be seen on \AT{2}, \AT{3} and
\AT{4}. It is clear that the combination of all observables will act as a way
to reduce the allowed regions for a model independent analysis.

\FIGURE{\includegraphics[width=0.49\textwidth]{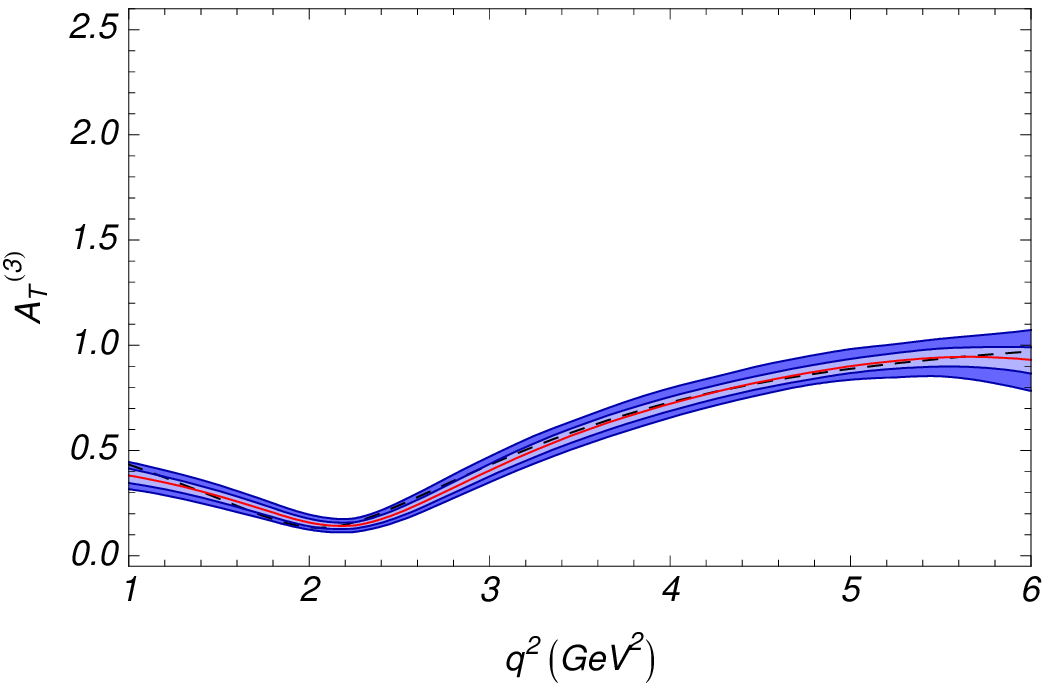}
  \includegraphics[width=0.49\textwidth]{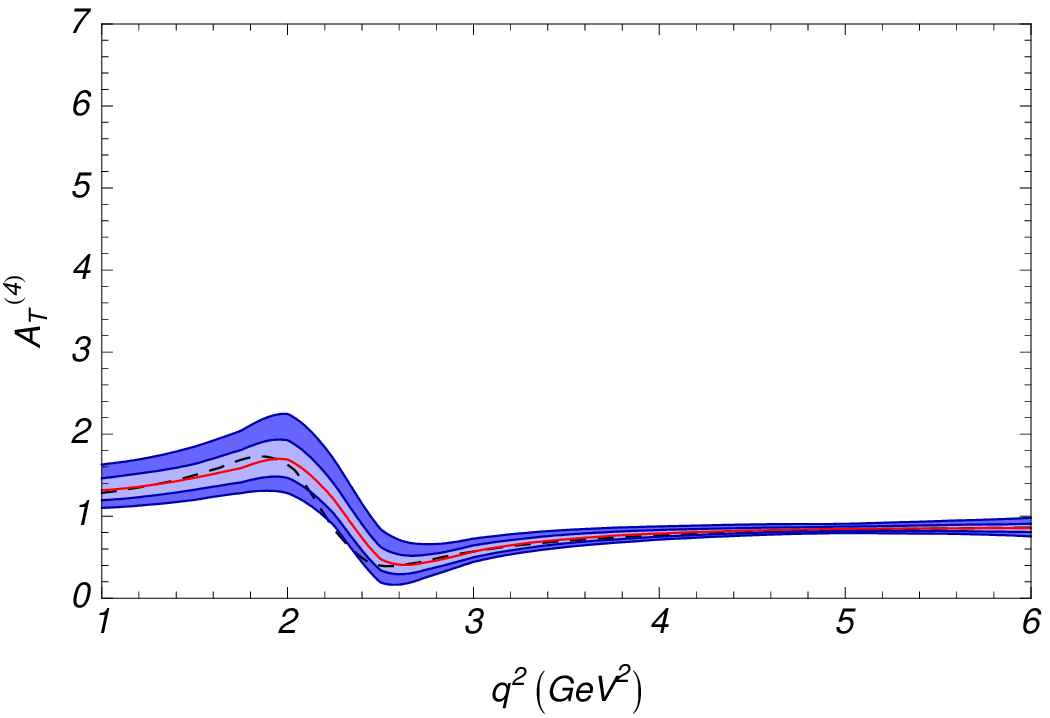}
  \caption{The experimental errors in \AT{3} (right) and \AT{4} left assuming
    the SM for statistics equivalent to 100\invfb at the end of an upgrade to
    \lhcb.}
  \label{fig:superLHCb}
}
Finally we might ask what happens if we consider the situation with 100\invfb
of experimental data corresponding to the full dataset from an upgrade to
\lhcb. We assume the same performance of the experiment so simply scale the
statistics by a factor 10 compared to the 10\invfb study. The experimental
errors are shown in Fig.~\ref{fig:superLHCb} and are in general just a factor
$\sqrt{10}$ smaller as expected. Comparing to Figs.~\ref{fig:AT3}
and~\ref{fig:AT4} it can be seen that the $\Lambda/m_b$ uncertainties will
dominate unless progress is made on the theoretical side.

\section{Summary}
\label{sec:summary}
We have constructed two new observables \AT{3} and \AT{4} out of the $K^*$
spin amplitudes of the \BdbKsmm decay, that fulfil the criteria of being
theoretically clean and can be experimentally extracted from the angular
distribution of this decay with good precision. We have shown how to design
the new observables for a specific kind of NP operator within the
model independent analysis using the effective field theory approach.

We have presented a complete calculation of all observables in QCD
factorization and have made the impact of unknown $\Lambda/m_b$ corrections to
the various observables explicit. Subsequently, we demonstrated the high
sensitivity of \AT{2}, \AT{3} and \AT{4} to right handed currents. Clearly
theoretical progress on the $\Lambda/m_b$ corrections would enhance that
sensitivity significantly and would be desirable in view of an
upgrade of the \lhcb experiment.

The new observables \AT{3} and \AT{4} exhibit the important property of
presenting a direct sensitivity to the longitudinal spin amplitude, while
reducing at maximum the sensitivity to the poorly known longitudinal soft form
factors within the whole low dilepton mass spectrum. Previously defined $F_L$
or \AFB does not exhibit this behaviour. This same idea was behind the
construction of \AT{2} using the transverse amplitudes.

The combination of the three observables offer a full view of the sensitivity
to NP of the three spin amplitudes with a good control of hadronic
uncertainties.

Using a toy Monte Carlo approach we have estimated the statistical uncertainty
of all observables for statistics corresponding to \lhcb and also for
Super-\lhcb. The model performs a fit to the full angular and \qsq
distribution. \AT{3} and \AT{4} require a full angular fit and for \AT{2} we
have demonstrated that the resolution improves by more than a factor 2
compared to extracting \AT{2} from angular projection. The experimental errors
are such that measuring these new observables will be a powerful way to detect
the presence of right handed currents. For the well known measurement of the
zero point of the forward-backward asymmetry we see an improvement of 30\% in
the resolution from a full angular fit compared to fitting the angular
projections.

Finally we have shown that the previously discussed angular distribution
\AT{1} cannot be measured at either \lhcb or at a Super-$B$ factory.

\acknowledgments
We thank Martin Beneke for detailed discussions on the forward-backward
asymmetry. JM acknowledges financial support from FPA2005-02211,
2005-SGR-00994 and the RyC programme, MR from the Universitat Autonoma de
Barcelona, and UE and WR from the Science and Technology Facilities
Council~(STFC). TH acknowledges support of the European network Heptools.

\newpage 

\appendix

\section{Kinematics}
\label{App:Kinematics}
Assuming the $K^*$ to be on the mass shell, the decay $\bar {B^0}\to \bar
K^{*0}(\to K^- \pi ^+) \ell^+ \ell^-$ is completely described by four
independent kinematic variables; namely, the lepton-pair invariant mass, \qsq,
and the three angles $\theta_l$, $\theta_{K^*}$, $\phi$ as illustrated in
\fig{fig:kinematics}.  The sign of the angles for the \Bdb decay shows great
variation in the literature. Therefore we present here the most explicit
definition of our conventions.
\FIGURE{\includegraphics[scale=0.9]{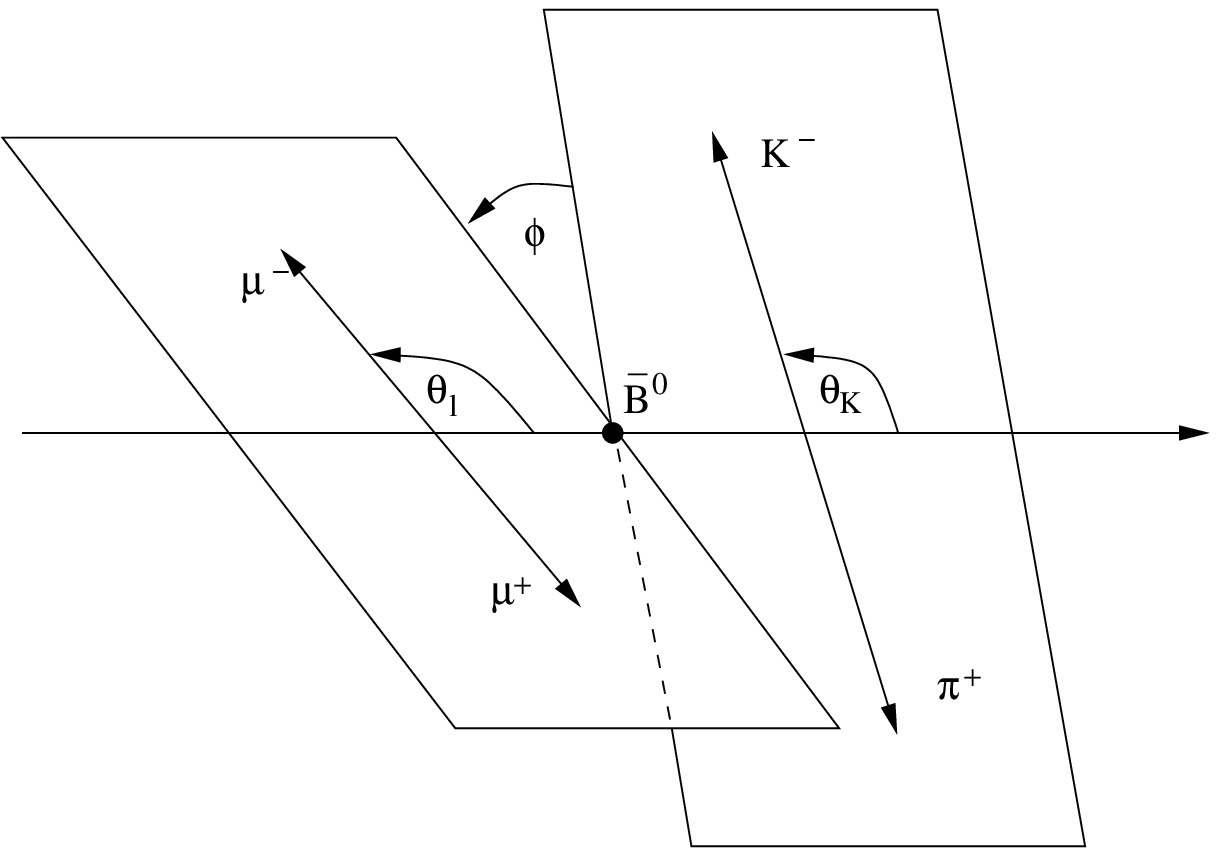}
  \caption{Definition of kinematic variables in the decay \BdbKsmm : The
    \mbox{$z$-axis} is the direction in which the \B meson flies in the rest
    frame of the \mumu. \thetaL is the angle between the \mun and the
    \mbox{$z$-axis} in the \mumu rest frame, \thetaK is the angle between the
    \Km and the \mbox{$z$-axis} in the \Kstarb rest frame, and $\phi$ is the
    angle between the normals to the \mumu and $\kaon\pi$ decay planes in the
    \B rest frame.}
  \label{fig:kinematics}
}
Here $\vect{p}$ denote three momentum vectors in the \Bdb rest frame,
$\vect{q}$ the same in the di-muon rest frame, and $\vect{r}$ in the \Kstarzb
rest frame, the $z$-axis is defined as as the direction of the \Kstarzb in the
\Bdb rest frame. Three unit vectors are given in the following way: the first 
one is in the direction of the $z$-axis
where the $\theta$ angles are measured with respect to, and the other two are
perpendicular to the di-muon and \Kstarzb decay planes.
\begin{equation}
  \label{eq:unitvecBdb}
  \vect{e}_z =  
  \frac{\vect{p}_{\Km}+\vect{p}_{\pip}}{|\vect{p}_{\Km}+\vect{p}_{\pip}|}\, ,
  \qquad
  \vect{e}_l=
  \frac{\vect{p}_{\mun}\times\vect{p}_{\mup}}
       {|\vect{p}_{\mun}\times\vect{p}_{\mup}|}\, , 
  \qquad 
  \vect{e}_K=
  \frac{\vect{p}_{\Km}\times\vect{p}_{\pip}}
       {|\vect{p}_{\Km}\times\vect{p}_{\pip}|}\, .
\end{equation}
It follows  for the \Bdb 
\begin{equation}
  \label{eq:AngleDefBdb}
  \cos\theta_l = \frac{\vect{q}_{\mun}\cdot\vect{e}_z}{|\vect{q}_{\mun}|}\, ,
  \qquad
  \cos\theta_{K} = \frac{\vect{r}_{\Km}\cdot\vect{e}_z}{|\vect{r}_{\Km}|}
\end{equation}
and
\begin{equation}
  \label{eq:angledef2Bdb}
  \sin\phi= (\vect{e}_l\times \vect{e}_K)\cdot \vect{e}_z\, , \qquad \cos\phi=
  \vect{e}_K\cdot \vect{e}_l\, .
\end{equation}

The angles are defined in the intervals
\begin{equation}
  \label{int:region}
  -1\leqslant\cos\theta_l\leqslant 1\, ,\qquad
  -1\leqslant\cos\theta_{K}\leqslant 1\, , \qquad
  -\pi\leqslant\phi < \pi\, ,
\end{equation}
where in particular it should be noted that the $\phi$ angle is signed.

In words, for the \Bdb the angle $\theta_l$ is measured as the angle between
the \mun and the $z$-axis in the dimuon rest frame. As the \Bdb flies in the
direction of the $z$-axis in the dimuon rest frame this is equivalent to
measuring $\theta_l$ as the angle between the muon and the \Bdb in the di-muon
rest frame. The angle $\theta_K$ is measured as the angle between the kaon and
the $z$-axis measured in the \Kstarzb rest frame.  Finally $\phi$ is the angle
between the normals to the planes defined by the $K\pi$ system and the
$\mu^+\mu^-$ system in the rest frame of the \Bdb meson.

\section{Theoretical framework}
\label{sec:theory}
The coefficient functions $I_i$ in the differential decay rate are given in
terms of the $K^*$ spin amplitudes (see Eq.~(\ref{eq:Isubis})) discussed in
Sec.~\ref{sec:amplitudes}.  The theoretical expressions of those spin
amplitudes can be derived using the following standard steps:

\begin{itemize}
\item 
The effective Hamiltonian describing the quark transition $b\to s
\ell^+\ell^-$ is given by
\be\label{heff}
{\mathcal{H}}_{\rm eff}=-\frac{4 G_F}{\sqrt{2}} V_{tb}^{} V_{ts}^*
\sum_{i=1}^{10}
[C_i (\mu) {\mathcal{O}}_{i}(\mu)  + C_i^\prime (\mu) {\mathcal{O}}_{i}^\prime
(\mu)],
\ee
where in addition to the SM operators we have also added  the chirally
flipped partners. 
In what follows, the same conventions are used as in~\cite{Kruger:2005ep}.
In the NP analysis, we will focus on  the  
the chirally flipped ${\cal O}_7^\prime$ operator in addition to
the most important SM operators ${\cal O}_7,{\cal O}_9,$ and ${\cal O}_{10}$:  
\begin{equation}\label{operator:basis}
{\mathcal{O}}_{7} = \frac{e}{16\pi^2} m_b
(\bar{s} \sigma_{\mu \nu} P_R b) F^{\mu \nu}\, , \qquad
{\mathcal{O}}_{7}^\prime = \frac{e}{16\pi^2} m_b
(\bar{s} \sigma_{\mu \nu} P_L b) F^{\mu \nu}\, ,
\end{equation}
\begin{equation}
{\mathcal{O}}_{9} = \frac{e^2}{16\pi^2}
(\bar{s} \gamma_{\mu} P_L b)(\bar{\ell} \gamma^\mu \ell)\, ,\qquad
{\mathcal{O}}_{10}=\frac{e^2}{16\pi^2}
(\bar{s}  \gamma_{\mu} P_L b)(  \bar{\ell} \gamma^\mu \gamma_5 \ell)\, ,
\end{equation}
where $P_{L,R}= (1\mp \g_5)/2$ and $m_b \equiv m_b(\mu)$ is the
running mass in the $\ol{\mbox{MS}}$ scheme.

\item The hadronic part of the matrix element describing the $B\to K\pi$
  transition can be parameterised in terms of $B\to K^*$ form factors by means
  of a narrow-width approximation (see for example \cite{FK:etal}).  The
  relevant form factors are defined as:
  \begin{eqnarray}\label{ff:btokstar:vector}
    \lefteqn{\braket{\kstar(p_{\kstar})}{\bar{s}\g_{\m}P_{L,R} b}{{B}(p)}
      =  i\epsilon_{\mu\nu\alpha\beta} \epsilon^{\nu*}p^{\alpha}q^{\beta} 
        \frac{V(\qsq)}{m_B+m_{\kstar}} \mp} \nnu \\
      & & \qquad \mp\frac{1}{2}\bigg\{\epsilon_{\m}^*(m_B+ m_{\kstar})A_1(\qsq) -(\epsilon^*\cdot q)(2p -q)_\mu \frac{A_2(\qsq)}{m_B+m_{\kstar}} - \nnu\\
    & & \qquad\qquad -
    \frac{2m_{\kstar}}{\qsq} (\epsilon^*\cdot q) [A_3(\qsq)- A_0(\qsq)]q_\mu\bigg\},
  \end{eqnarray}
  where 
  \begin{equation}
    A_3 (\qsq) = \frac{m_B+m_{\kstar}}{2m_{\kstar}} A_1(\qsq) - \frac{m_B-m_{\kstar}}{2m_{\kstar}}A_2(\qsq)\, ,
  \end{equation}
  and 
  \begin{eqnarray}\label{ff:btokstar:tensor}
    \lefteqn{\braket{\kstar(p_{\kstar})}{\bar{s}i \sigma_{\mu\nu}q^{\nu}P_{R,L} b}{{B}(p)}
      =  - i\epsilon_{\mu\nu\alpha\beta} \epsilon^{\nu*}p^{\alpha}q^{\beta}T_1(\qsq)
      \pm} \nnu  \\
     & & \qquad \pm \frac{1}{2}\bigg\{[\epsilon_{\mu}^*(m_B^2-m_{\kstar}^2) - (\epsilon^*\cdot q)(2p-q)_{\mu}] T_2(\qsq) + \nnu \\
    & & \qquad\qquad + (\epsilon^*\cdot q)
    \bigg[ q_\mu - \frac{\qsq}{m_B^2-m_{\kstar}^2} (2p - q)_\mu\bigg]T_3(\qsq)\bigg\}.
  \end{eqnarray}
  In the above, $q = p_{l^+}+ p_{l^-}$ and $\epsilon^{\m}$ is the $K^*$
  polarisation vector.

\item In the heavy-quark and large-energy limit the seven a priori independent
  $B\to K^*$ form factors in Eqs.~(\ref{ff:btokstar:vector}) and
  (\ref{ff:btokstar:tensor}) reduce to two universal form factors $\xi_{\bot}$
  and $\xi_{\|}$ in the leading order
  \cite{large:energy:limit,Beneke:2001wa}:\footnote{Following
    \rf{Beneke:2001wa}, the longitudinal form factor $\xi_\|$ is related to
    that of \rf{large:energy:limit} by $\xi_{\|} = (m_\kstar/E_\kstar)
    \zeta_{\|}$.}
  \begin{subequations}\label{form:factor:relations:LEL}
    \begin{equation}
      A_1(\qsq)= \frac{2 E_\kstar}{m_B + m_\kstar}\xi_{\bot}(E_\kstar)\, ,
    \end{equation}
    \begin{equation}
      A_2(\qsq)= \frac{m_B}{m_B-m_\kstar}
      \bigg[\xi_{\bot}(E_\kstar)- \xi_{\|}(E_\kstar)\bigg],
    \end{equation}
    \begin{equation}
      A_0(\qsq) = \frac{E_\kstar}{m_\kstar} \xi_{\|}(E_\kstar)\, ,
    \end{equation}
    \begin{equation}
      V(\qsq)= \frac{m_B+m_\kstar}{m_B}\xi_{\bot}(E_\kstar)\, ,
    \end{equation}
    \begin{equation}
      T_1(\qsq)=\xi_{\bot}(E_\kstar)\, ,
    \end{equation}
    \begin{equation}
      T_2(\qsq)=\frac{2 E_\kstar}{m_B}\xi_{\bot}(E_\kstar)\, ,
    \end{equation}
    \begin{equation}
      T_3(\qsq)=\xi_{\bot}(E_\kstar) - \xi_{\|}(E_\kstar)\, .
    \end{equation}
  \end{subequations}%
  Here, $E_\kstar$ is the energy of the final vector meson in the ${B}$ rest
  frame,
  \begin{equation}
    E_\kstar \simeq  \frac{m_B}{2}\left(1-\frac{\qsq}{m_B^2} \right).
  \end{equation}
  These relations, valid in the low-\qsq region, allow to simplify the spin
  amplitudes to obtain
  Eqs.~(\ref{LEL:tranversity:perp}--\ref{LEL:tranversity:zero}) which are
  crucial for the construction of our new observables.  They are violated by
  symmetry breaking corrections of order $\alpha_s$ and $1/m_b$.
\end{itemize}

\section{NLO corrections to the spin  amplitudes}
\label{sec:NLO}
The NLO corrections to the form factors at order $\alpha_s$ are given in
Ref.~\cite{Beneke:2001at}. In the presence of right-handed currents
($\Cpeff{7}\neq 0$), the spin amplitudes read
\cite{Kruger:2005ep,Lunghi:2006hc}
\begin{equation}
  \label{a_perpNLL}
A_{\bot L,R}=N \sqrt{2}\, \lambda^{1/2}\bigg[
(\C{9}\mp\C{10})\frac{V(\qsq)}{m_B +m_\kstar}+\frac{2m_b}{\qsq}
{\cal T}_{\perp \rm NLO}^+(\qsq)\bigg],
\end{equation}
\begin{equation}
  \label{a_parNLL}
  A_{\| L,R}= - N \sqrt{2}\,\,(m_B^2- m_\kstar^2)\bigg[(\C{9} \mp \C{10})
  \frac{A_1 (\qsq)}{m_B-m_\kstar}
  +\frac{4 m_b \, E_{K^*}}{m_B \cdot s} {\cal T}_{\perp \rm NLO}^-(\qsq)\bigg],
\end{equation}
\begin{eqnarray}\label{a_longNLL}
  A_{0L,R}&=&-\frac{N}{2m_\kstar q}\times \nnu \\ 
  & & \times \bigg[
  (\C{9}\mp \C{10})\bigg\{(m_B^2-m_\kstar^2 -\qsq)(m_B+m_\kstar)A_1(\qsq)
  -\lambda \frac{A_2(\qsq)}{m_B +m_\kstar}\bigg\}\nnu\\
  & &\quad\  {2m_b} \bigg\{
  (m_B^2+3m_\kstar^2 -\qsq)\frac{ 2 E_{K^*}}{m_B}{\cal T}_{\perp \rm NLO}^-(\qsq)
  \\ 
  & & \qquad \qquad -\frac{\lambda}{m_B^2-m_\kstar^2}\left({\cal T}_{\perp \rm NLO}^-(\qsq) + {\cal T}_{\| \rm NLO}^-(\qsq)
  \right)  \bigg\}\bigg], \nonumber
\end{eqnarray}
where $\lambda$ is defined in Eq.~\ref{eq:Lambdadef} and the form factor relations for $V(\qsq), A_0(\qsq)$ and $A_1(\qsq)$ are as in  
Eq.~\ref{LEL:tranversity:perp}. $A_2(\qsq)$ is given by 
\begin{equation}
  A_2(\qsq)= \frac{m_B}{m_B-m_\kstar}
  \left[\xi_{\bot}(\qsq)- \xi_{\|}(\qsq)\left(1-C\right)\right],
\end{equation}
with $C$ the $\mathcal{O}(\alpha_s)$ correction to the form factor $A_2$
computed in \cite{Beneke:2001wa,Beneke:2001at,Beneke:2004dp}:
\begin{equation}
  C= \frac{\alpha_s}{3\pi}\left[\left(2-2L\right)+ 8\, \frac{m_\kstar}{E_\kstar} \, 
    \frac{m_B(m_B-2E_\kstar)}{4E_\kstar^2}\,\kappa_{\|}(\qsq) \,  \lambda_{B,+}^{-1} \int_0^1 du \frac{\Phi_{\kstar,\|}(u)}{1-u} \right]
\end{equation}
with $\Phi_{K^*,\|} (u)$ being the longitudinal light-cone distribution amplitude of the vector meson $\Kstarzb$. Moreover, we have  
\begin{eqnarray}\label{nlodef}
  {\cal T}_{\perp \rm NLO}^{\pm} &=& \xi_\perp(\qsq)
  \Bigg\{ C_\perp^{(0,\pm)}  +
  \frac{\alpha_s}{3\pi} \Bigg[
  C_\perp^{(1,\pm)} + \nnu \\
  &  & + \kappa_\perp(\qsq)\,
  \lambda^{-1}_{B,+}\int_0^1 du\,\Phi_{K^*, \perp}(u)\left[T_{\perp,+}^{(\mathrm{f}\pm)}(u)+
    T_{\perp,+}^{(\mathrm{nf})}(u)\right] \Bigg] \Bigg\}
  \nonumber
\end{eqnarray}
and
\begin{eqnarray}
  \lefteqn{{\cal T}_{\parallel \rm NLO}^\pm =  \xi_\parallel(\qsq) \left\{
    C_\parallel^{(0, \pm)}
    +\kappa_\|(\qsq) \,\frac{m_K^*}{E_\kstar} \,
    \lambda^{-1}_{B,-}(\qsq)
    \int_0^1 \!du\,\Phi_{K^*,\,\parallel}(u)
    \,\hat{T}_{\parallel,\,-}^{(0)}(u) \right. +}
  \nonumber\\
  & & + \left. \frac{\alpha_s}{3\pi} \left[
      C_\parallel^{(1, \pm)}  + 
      \kappa_\|(\qsq) \frac{m_K^*}{E_\kstar} \left( \,
        \lambda^{-1}_{B,+}
        \int_0^1 \!du\,\Phi_{K^*,\,\parallel}(u)
        \,\left[T_{\parallel,\,+}^{(\mathrm{f} \pm)}(u)+
          T_{\parallel,\,+}^{(\mathrm{nf})}(u)\right] + \right. \right. \right. \nonumber \\
  &  & + \left.\left.\left.
        \lambda^{-1}_{B,-}(\qsq)
        \int_0^1 \!du\,\Phi_{K^*,\,\parallel}(u)
        \,\hat{T}_{\parallel,\,-}^{(\mathrm{nf})}(u) \right) \right] \right\} \label{taus},
\end{eqnarray}
where $\kappa_z\equiv\pi^2 f_B f_{K^*,\,z}(\mu)/{(N_c m_B \xi_z(\qsq))}$ (with
$z=\perp,\|$).
$\lambda_{B,+}^{-1}$ and $\lambda_{B,-}^{-1}(\qsq)$ are the two $\Bdb$ meson light-cone distribution amplitude moments defined in~\cite{Beneke:2001at}; they are given by
\begin{subequations}
\begin{eqnarray}
 \lambda_{B,+}^{-1} &=& \int_0^\infty d\omega \frac{\Phi_{B,+}(\omega)}{\omega}\,, \\
 \lambda_{B,-}^{-1} (\qsq) &=& \int_0^\infty d\omega \frac{\Phi_{B,-}(\omega)}{\omega-\qsq/m_B -i\epsilon}\,.
\end{eqnarray}
\end{subequations}
In all cases the symbol $+$ stands for the substitution of $\Ceff{7} \to
\Ceff{7}+\Cpeff{7}$ and $-$ for $\Ceff{7} \to \Ceff{7}-\Cpeff{7}$, wherever
$\Ceff{7}$ appears.  For instance, in the definition of 
\begin{equation}
 C^{(1,\pm)}_{z} =  C_{z}^{(\rm f\pm)}+ C_{z}^{(\rm nf)} 
\end{equation}
with $z=\perp,\|$, the factorizable
correction reads~\cite{Beneke:2001wa,Beneke:2001at}
\begin{eqnarray}
  \label{start2}
  && \hspace*{-0.5cm}
  C^{(\rm f\pm)}_{\perp} = \left(C_7^{\,\rm eff}\pm C_7^{\,\rm eff \prime}\right)\left(4\ln\frac{m_b^2}{\mu^2}
    -4-L \right),\\
  && \hspace*{-0.5cm}
  C^{(\rm f\pm)}_{\parallel} = - \left(C_7^{\,\rm eff}\pm C_7^{\,\rm eff \prime}\right) \left(4\ln\frac{m_b^2}{\mu^2}
    -6+4 L\right)
  + \frac{m_B}{2 m_b}\,Y(\qsq)\,\Big(2-2 L\Big)
  \label{end2}
\end{eqnarray}
with
\begin{equation}
\label{Ldef}
  L\equiv -\frac{m_b^2-\qsq}{\qsq}\ln\left(1-\frac{\qsq}{m_b^2}\right).
\end{equation}
while the non-factorizable contribution $C_{z}^{(\rm nf)}$ is common to both.
In the definition of the hard scattering functions with $T^{(1 \pm)}_{z,\,\pm}
= T_{z,\,\pm}^{(\rm f\pm)}+ T_{z,\,\pm}^{(\rm nf)}$, the
factorizable correction reads \cite{Beneke:2001wa,Beneke:2001at}:
\begin{eqnarray}
  \label{start1}
  && \hspace*{-0.5cm}
  T^{(\rm f\pm)}_{\perp,\,+}(u,\omega) = \left(C_7^{\,\rm eff}\pm C_7^{\,\rm eff \prime}\right) \,
  \frac{2 \, m_B}{(1-u) E_\kstar}\, ,\\
  && \hspace*{-0.5cm}
  T^{(\rm f\pm)}_{\parallel,\,+}(u,\omega) = \left[\left(C_7^{\,\rm eff}\pm C_7^{\,\rm eff \prime}\right)  +
    \frac{\qsq}{2 \, m_b \, m_B} \,Y(\qsq)\right]\,
  \frac{2 \, m_B^2}{(1-u) E_\kstar^2}\, ,
  \\
  && \hspace*{-0.5cm}
  T^{(\rm f)}_{\perp,\,-}(u,\omega) =
  T^{(\rm f)}_{\parallel,\,-}(u,\omega) = 0\, .
\end{eqnarray}
Again the non-factorizable part is common to both cases, because it does not
receive contributions from \Ope{7}.  For the definition of the function
$Y(\qsq)$ and for the non-factorizable contributions we refer the reader to
\cite{Beneke:2001wa,Beneke:2001at}.

\end{document}